\documentclass[]{interact}
\usepackage{epstopdf}
\usepackage[caption=false]{subfig}

\usepackage[numbers,sort&compress]{natbib}
\bibpunct[, ]{[}{]}{,}{n}{,}{,}

\theoremstyle{plain}

\newtheorem*{question}{Question}
\newtheorem*{answer}{Answer}

\theoremstyle{definition}

\theoremstyle{remark}

\newcommand\apj{\emph{The Astrophysical Journal}}%
\newcommand\apjl{\emph{The Astrophysical Journal Letters}}
\newcommand\mnras{\emph{Monthly Notices of the Royal Astronomical Society}}%
\newcommand\araa{\emph{Annual Review of Astronomy and Astrophysics}}
\newcommand\aaps{\emph{Astronomy and Astrophysics Supplements}}
\usepackage{amsmath}
\usepackage{graphicx,psfrag,epsf}
\usepackage{enumerate}
\usepackage{natbib}
\usepackage{url} 
\usepackage{amssymb,amsmath}
\usepackage{color,xcolor,nicefrac}
\def\fuv{\emph{FUV}}
\def\pg{PG~1116+215}

\def\rgs{\emph{RGS}}
\def\nodata{---}

\def\oi{O~I}

\def\ovii{O~VII}
\def\oviii{O~VIII}

\def\chandra{\it Chandra\rm}
\def\xmm{\emph{XMM-Newton}}

\def\letg{\emph{LETG}}
\def\blue{\color{black}}
\begin{document}

\articletype{APPLICATION NOTE}

\title{Probability models of chance fluctuations in spectra of astronomical sources
{\blue with applications to X--ray absorption lines}}

\author{
\name{Massimiliano Bonamente\textsuperscript{a}\thanks{
The author gratefully acknowledges support of \textit{NASA} 
\chandra\ grant AR6-17018X.}}
\affil{\textsuperscript{a} Department of Physics and Astronomy,
University of Alabama in Huntsville, Huntsville AL 35899 (U.S.A)}}
\maketitle

\begin{abstract}
The search for faint emission or absorption lines
in astronomical spectra
has received considerable attention in recent years,
{\blue especially in the X--ray wavelength range}.
These features usually appear as a deficit or excess of
counts in a single resolution element of the
detector, and as such they are referred to as \emph{unresolved} fluctuations.

The general problem under
investigation is the probability of occurrence 
of chance fluctuations.
A quantitative answer is necessary to determine whether
detected fluctuations are a real (astronomical, in this case) signal, 
or if they can be attributed to chance.
This application note provides a new comprehensive method to
answer this question as function of the instrument's resolution, 
the wavelength coverage of the spectrum, the number of fluctuations of interest,
 and the confidence level chosen.

The method is based on the binomial distribution, 
and addresses
also the probability of multiple simultaneous fluctuations.
A critical aspect of the model is the \emph{a priori} knowledge of the
location of possible fluctuations, which significantly affects the probability of
detection. In fact, a wider wavelength range for the `blind' search of possible fluctuations
results in a larger number of `tries' for the detection of a fluctuation, 
lowering the overall significance of a specific feature. 
The method is illustrated with
a case study and examples using X--ray data.

\end{abstract}


\section{Introduction}
\label{sec:introduction}

The wavelength distribution of radiation from astronomical sources, commonly
referred to as \emph{spectrum}, is a fundamental tool  for astronomy.
In particular, it is common to search for emission or absorption features 
as a fluctuation above or below a well--characterized baseline emission level,
or \emph{continuum},
in the spectrum. Such
features may originate from a number of astrophysical phenomena, 
such as emission or absorption lines, and carry valuable information on the 
emitting source or
the intervening inter--stellar and inter--galactic medium between the source 
and the observer.
Often these features are sufficiently weak that it is necessary to study whether
they may be simply a random fluctuation due to noise in the spectrum, rather
than a real feature of astronomical origin. 

This paper focuses on a new comprehensive
method to establish when one or multiple spectral features are statistically
significant. 
{\blue Although the methods presented in this paper can be applied to any spectrum,
examples and applications are from X--ray absorption lines in the spectra of
distant astronomical sources that have recently appeared in the literature
\cite{bonamente2016,bonamente2017b,nicastro2018}.}
 This introductory section first provides a review of
basic concepts of spectroscopy, then a review of current methods and
the need for a new approach are presented.

\subsection{Spectra of astronomical sources}
\label{sec:introduction-1}
\begin{figure}[!t]
\centering
\includegraphics[width=5in,angle=-0]{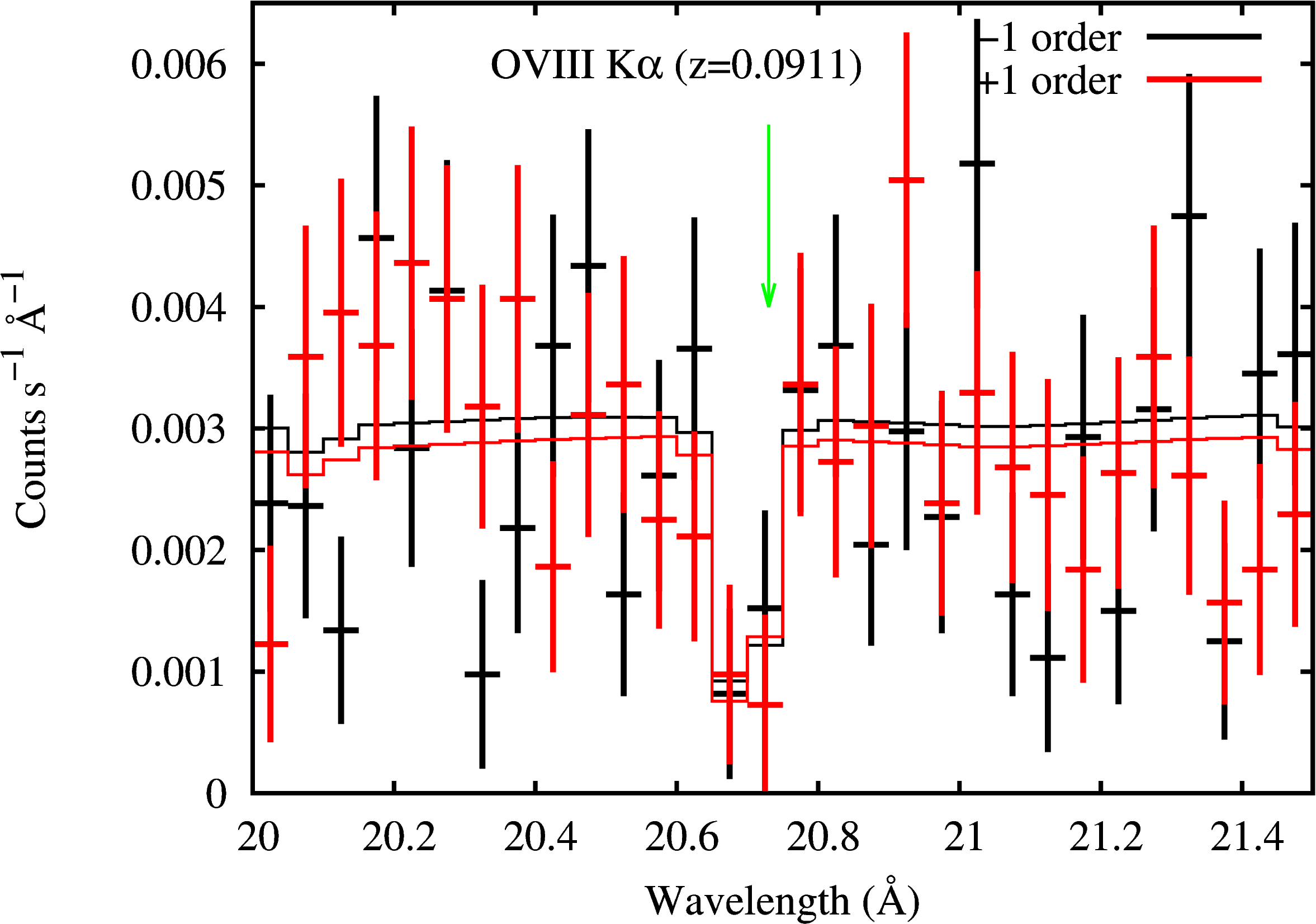}
\caption{\small A portion of the \chandra\ X--ray spectrum of 
\pg, from \cite{bonamente2016}. Two sets of data points (red and black)
correspond to two spectra collected by the same instrument, caused by
incoming photons being dispersed in opposite directions on the detector
($\pm 1$ order). The line was identified by \cite{bonamente2016} as 
a K$\alpha$ line from \oviii, red-shifted by $z=0.0911$.
Green arrow marks the position of the $z=0.0928$ \oviii\ K$\alpha$ line,
discussed in detail in Section~\ref{sec:pg1116}.}
\label{fig:zoom}
\end{figure}

This paper considers \emph{unresolved} features, defined as fluctuations with a
wavelength extent that is smaller than the resolution of the instrument.
The resolution of the instrument is characterized by a \emph{full--width at half maximum}
(FWHM) of the response to a narrow (or impulse) function. For example,
current--generation X--ray instruments such as the \rgs\ on board \xmm\ and the \letg\ 
on board \chandra\ have a resolution of order $\sigma_{\lambda} \sim 
50$~m\AA50~m\AA\footnote{\AA\ indicates Angstrom, 
a unit of measure of distance equivalent to $10^{-10}$~m.}, meaning
that an intrinsically narrow spectral feature will be spread over a wavelength range of about  
50~m\AA, 
purely by instrumental effects.
An unresolved feature is intended to be confined to approximately one 
or two resolution elements, depending on binning of the data.
Figure~\ref{fig:zoom} shows a portion of the X--ray spectrum of an extragalactic
source (\pg), obtained with the \chandra\ satellite. The spectrum is binned to a
size of 50~m\AA, approximately the size of the resolution
of the instrument. The apparent absorption feature near 20.7~\AA\ is 
the type of fluctuations of interest, and it will
discussed in more
detail as part of the case study of Section~\ref{sec:pg1116}.
{\blue The aim of this paper is to provide a statistical method to determine 
whether such features are a real (astronomical) signal, or whether
they are due to random fluctuations in the spectrum. }

An example of spectral lines is given by
the \ovii\ K$\alpha$ absorption line from the interstellar or intergalactic medium, 
often occurring in the
spectrum of a background quasar \citep[e.g.][]{bregman2007}. 
The rest wavelength of this line is $\lambda_0=21.60$~\AA, 
and the \ovii\ ion (i.e., a \emph{six}--times ionized oxygen atom) is abundant at temperatures of order
$ T= 10^6$~K, characteristic of a certain phase of the
interstellar and intergalactic media \citep{mazzotta1998, verner1996}. 
Another example is the \oviii\ K$\alpha$ absorption line, with rest wavelength 
$\lambda_0=18.97$~\AA, abundant at higher temperatures than \ovii. 
Much attention has been given in the recent literature
to the detection of such absorption lines using X--ray spectra \citep[e.g.][]{fang2002,fang2005,williams2007,
nicastro2005,nicastro2013,bonamente2016}.
The spectrum of Figure~\ref{fig:zoom} is an excerpt from \cite{bonamente2016}, where
a detection of \oviii\ was claimed. In Section~\ref{sec:N} is explained how
a line from an astronomical source, such as that of Figure~\ref{fig:zoom},
 can be shifted from its rest wavelength.

Spectral lines are not infinitely sharp impulse functions. Lines
are always broadened by a number of atomic and astrophysical phenomena
that provide the line with a specific line profile, even without accounting
for the broadening due to the instrumental resolution.
A typical source of broadening is the temperature of the
plasma responsible for the line, with a typical thermal velocity dispersion for \ovii\ lines
of order $50-100$~km~s$^{-1}$. {\blue Such velocity dispersion
is simply given by the fact that plasma particles at a given temperature have velocities
given by a Maxwellian distribution.}
The relationship 

\[ \lambda = \lambda_0 \left( 1 + \frac{v}{c} \right) \]

describes the Doppler wavelength shift caused by a radial velocity $v$, 
assumed non--relativistic ($v \ll c$, where $c$ is the speed of light).
A typical radial velocity $v = 100$~km~s$^{-1}$ corresponds
to a wavelength dispersion $\Delta \lambda \simeq 7$~m\AA\ around the nominal 
\ovii\ K$\alpha$ wavelength $\lambda_0 = 21.6$~\AA. 
Therefore the line profile cannot be resolved by the available X--ray instruments
with a resolution of $\sigma_{\lambda} \sim 50$~m\AA, 
and the fluctuation will appear as a deficit of counts
in one or perhaps two (depending on binning of data)
 resolution elements of extent $\sigma_{\lambda}$.
 This situation is typical of
X-ray emission and absorption lines in the spectra of bright X--ray sources such as quasars.
Different considerations would be used
when analyzing \emph{resolved} features, i.e., those that can be detected in several resolution
elements, as is typically the case for visible or far--ultraviolet (FUV) lines detected
by instruments with better spectral resolution \citep[e.g.,][]{tilton2012}.
{\blue Such extended and resolved features are not considered in this paper.}

\subsection{Current methods to estimate the significance of fluctuations in astronomy
and their challenges}
\label{sec:literature}

The significance of detection of unresolved fluctuations, e.g., faint absorption or emission
lines, has received considerable attention in astrophysics in recent years.
In particular this is true for the emerging field of X--ray
spectroscopy, where
the detectors have a much lower spectral resolution than the more traditional
and developed optical or infra--red astronomy fields. Such emission or absorption lines
will appear as fluctuations in a single resolution element, thus making the discussion of
Section~\ref{sec:introduction-1} relevant.
This section describes current methods to assess the significance of
faint and unresolved
absorption lines in X--ray astronomy. The aim is to highlight the progress in
 statistical methods in this field
and the need for a more comprehensive framework, which will
be described in Sections~\ref{sec:problem}--\ref{sec:model2}.

The standard method to assess the significance of a fluctuation in the spectrum
of an astronomical source is that of simply determining the signal--to--noise
ratio of the fluctuation, relative to a suitable reference level. In the case
of a fluctuation described by a Gaussian statistic, i.e., in the high--count
approximation of the Poisson counting statistic, the significance 
or signal--to--noise ratio

\[ S/N  = \left| \frac{\text{measured fluctuation}}{\text{error in the measured fluctuation}} \right| \]

measures the number of standard deviations of the fluctuation from the reference level.
For example, a value of $S/N=3$ implies a null hypothesis probability of 0.3\% of finding such a
deviation by chance. This probability is used to determine whether 
a feature is real or not. This is the method used, for example, by
\cite{nicastro2005}, and it is accurate only if one has an 
\emph{a priori} reason to expect a fluctuation in the specific
spectral bin (i.e., at the wavelength) where it was detected.

Random fluctuation can however occur at \emph{any} of the bins available in the spectrum.
To improve on the earlier method, \cite{nicastro2013} explicitly introduced the
concept of `\emph{number of redshift trials}'; that is to say, 
to account for the fact that there are many resolution elements in the spectrum where
the feature could appear. In Section~\ref{sec:approximation} it will be shown that the 
method of \cite{nicastro2013} is in fact an approximation to the more general
method that will be presented in the following sections. 
In \cite{fang2002} it is also discussed the need to account for the number of possible resolution
elements, although there was no detailed explanation of
their method to assess the statistical significance of
the absorption lines. 

A method to establish the joint significance of several absorption
line features was also discussed in \cite{nicastro2005}. Having tentatively
identified a set of negative fluctuations at specific wavelengths as possible
absorption lines from various elements, the authors use an $F$--test to
determine whether the additional model that describes the intensity of
the absorption lines is significant. This method is accurate only
under the conditions that all fluctuations are expected \emph{ a priori} at
the wavelengths at which they were detected.

The main challenge in the estimation of the statistical significance of
fluctuations is the lack of a general framework with a rigorous statistical
and probabilistic foundation.
The aim of this paper is to provide such general framework to
study the statistical significance of one or several fluctuations
(e.g., unresolved absorption lines), whether they
appear at  predetermined wavelengths, or randomly (i.e., during
a blind search). Such comprehensive framework 
has not appeared thus far in the refereed literature. 
It will be shown that some of the methods
presented earlier in the literature, such as the number of redshift
trials of \cite{nicastro2013}, can be obtained as an approximation of the methods
presented in this paper.

\section{Methodology I: Statement of the statistical problem}
\label{sec:problem}
The general question of interest is that of how many fluctuations
{\blue of a given significance} are expected in 
a spectrum purely because of
random errors in the measurements. This section presents first a 
definition of fluctuations in the 
context of astronomical spectra and the associated probability models.
Then a discussion follows on how \emph{priors} can be used to
reduce the portion of wavelength space that can be searched for
such fluctuations. These two aspects are then incorporated in the
statistical models presented in 
Sections~\ref{sec:model1}--\ref{sec:model2}.

\subsection{Definition of fluctuations}
\label{sec:DeltaK}
A fluctuation can be defined as the deviation 
of detected counts, or an equivalent quantity such as flux,
from a true underlying model of the source,

\begin{equation}
 \Delta K = K - K_{\text{model}}
\label{eq:DeltaK}
\end{equation}

where $K$ is the number of counts detected in the resolution element, and $K_{\text{model}}$ is the
number of counts predicted by the model in that bin, 
e.g., from a fit to the continuum of the astronomical source
{\blue in a broader wavelength range}.
The statistical significance of the number $\Delta K$ depends on the
properties of both random variables $K$ and $K_{\text{model}}$. 
The null hypothesis is that the random variable (or measurement) $K$ is drawn from the
same distribution as $K_{\text{model}}$, i.e., that the 
underlying model is an accurate description of the measurement $K$.
As such, the expectation or mean of $\Delta K$
must be zero, under the null hypothesis.
The probability distribution function of 
$\Delta K$ (e.g., Gaussian or other) and therefore its variance
and confidence intervals depend on the nature of the fluctuations. 

Before probability models
for $\Delta K$ are discussed in the following section, a  
note on Equation~\ref{eq:DeltaK}. Although the equation
describes the fluctuation of a given spectral bin, it 
does incorporate information from the entire spectrum,
since all bins are used to 
fit the continuum and thus determine the probability distribution
of $K_{model}$. For example, if the model of the continuum is a power--law,
uncertainties in the parameters of this model (normalization and index)
depend on the statistical fluctuation of all data point used in the 
fit, with larger fluctuations among data points leading to larger errors
in $K_{model}$.
As a result, information about all other spectral bins 
and their fluctuations are in fact 
incorporated in the assessment of $\Delta K$ for a specific
bin.

\subsection{Probability models for $\Delta K$ fluctuations}
\label{sec:probabilityModels}
A general reference for the modeling of the random variable $\Delta K$ 
is \cite{bonamente2017book}, or any other  textbook on the general theory of probability 
\citep[e.g.][]{ross2012}.
Recall the definition of a fluctuation as the variable $\Delta K = K - K_{\text{model}}$, where

\[ \begin{cases} 
K: \text{Number of counts or flux \emph{detected} in a resolution element} \\
K_{\text{model}}: \text{Number of counts or flux \emph{predicted} in a resolution element.}
\end{cases}
\]

Given that the flux is typically in units of counts (or energy) per unit time and area, 
and that area and time are known exactly, the properties of $K$ are those
of a counting variable. In most cases this means that $K$ is Poisson distributed.
When the number of counts exceeds approximately 20, the Poisson distribution
is accurately approximated by a Gaussian of 
same mean and standard deviation
\citep[see, e.g., chapter 3.4 of][]{bonamente2017book}. A central confidence interval
for either a Gaussian or Poisson variable
can be constructed using Tables~\ref{tab:K-Gaussian} and \ref{tab:K-Poisson}, where
$K=n$ is the number of detected counts.
In particular, confidence intervals for Poisson variables are obtained using the
Gehrels approximation \citep{gehrels1986}, also described in \cite{bonamente2017book}.
For example, a 68.3\% confidence interval for a Gaussian variable is
a range of $\pm 1 \sigma$ around its mean, and  a 68.3\% confidence interval on 
for Poisson variable with $n=10$ is the range from 6.90 to 14.28,
using the $S=1$ case of Table~\ref{tab:K-Poisson}.
The Poisson parameter $S$ is used to relate Gaussian confidence intervals
to the non--symmetric Poisson confidence intervals, 
as explained in \cite{bonamente2017book}.

\begin{table}[!t]
\centering
\caption{Confidence intervals for Gaussian variables}
\label{tab:K-Gaussian}
\begin{tabular}{lll}
\hline
Gaussian Interval & Enclosed Probability &  Poisson $S$ \\
\hline
$\pm 0.68 \sigma$ & 0.50	& 0.68	\\
$\pm 1 \sigma$    & 0.683	& 1 	\\
$\pm 1.65 \sigma$ & 0.90	& 1.65	\\ 
$\pm 2 \sigma$    & 0.955	& 2 	\\
$\pm 3 \sigma$    & 0.997	& 3 	\\
$\pm 4 \sigma$    & 0.99994	& 4	\\
\hline
\end{tabular}
\end{table}

The continuum model is typically obtained from a fit to the spectrum over a large wavelength
range.
The variance of $K_{model}$ is often much smaller than that of the fluctuation $K$,
because of the wider baseline of data used for its determination.
When the model is known accurately, the probability model of $\Delta K$ is just that of
$K$. In this case Tables~\ref{tab:K-Gaussian} and \ref{tab:K-Poisson} are sufficient
to model the fluctuation. Another case that is easily treatable is that of
Gaussian $K$ and Gaussian $K_{model}$; assuming independence between the
two variables, $\Delta K$ remains
Gaussian with a variance equal to the sum of the two variances.

In astronomy the fluctuation $\Delta K$ is often modeled directly during the
spectral fit using available software. A characteristic example
is described in \cite{bonamente2016}, where
the continuum is modeled via a power--law component
and the absorption line as a Gaussian~\footnote{The profile
can be alternatively described by a Lorentz or Voigt profile, and all
conclusions will apply to any profile used. The choice of Gaussian parameterization
for the line profile is completely unrelated to the Gaussian distribution
for any of the model parameters.} profile with known central wavelength and width,
and with variable depth. In this case, the depth of the Gaussian profile is the  parameter
that describes 
the fluctuation $\Delta K$, since it is proportional to the
number of counts in the absorption line. 
The  probability distribution function of $\Delta K$ is generally unavailable,
except for the simple cases described earlier in this section. The modelling of the
difference between counting variables has also received attention in other fields,
such as econometrics \cite{cameron2004}. One of the challenges of finding
the probability distribution function of $\Delta K$ is the fact that the joint
distribution of $K$ and $K_{model}$ is unknown (as is the correlation between
them). Therefore it is in general not possible to know
exactly the distribution function of $\Delta K$. 
This limitation can be circumvented by using an approach
that establishes the confidence interval for $\Delta K$, without knowing its
probability distribution function. This method is described in the following.

The confidence level of $\Delta K$ is calculated 
by the analysis software using
the $\chi_{min}^2$ statistic (or, alternatively, the Cash statistic $\mathcal{C}$,
if $K$ is in the low--count regime),
by varying the parameter of interest until $\chi^2$ or $\mathcal{C}$ is increased from its minimum
by a prescribed amount (e.g., by $+1$ for a 68\% confidence interval on the parameter,
or $+2.7$ for a 90\% confidence interval, as explained in \cite{bonamente2017book}).
During this procedure the other parameters, in particular those describing the continuum,
are also adjusted, and the result is a confidence level on the parameter of interest,
in this case $\Delta K$,
that accounts for variability of the other parameters. 
This $\Delta \chi^2$ method of analysis is the standard method used in X--ray 
spectral analysis \citep[e.g.][]{lampton1976}, and it provides a confidence interval
for $\Delta K$ at the desired probability level.
The method does away with the need to study $K$ and $K_{model}$ separately, and it
provides a convenient way to describe the statistical properties of $\Delta K$ directly.
In general, $\Delta K$ is not Gaussian distributed, and one needs not know its distribution.
The method however does provide a confidence interval (e.g., at 68\%, 90\% or any other level
of probability), and this information is sufficient to make an assessment of
its statistical significance.  

\begin{table}[!t]
\centering
\caption{Confidence intervals for Poisson variables. The Poisson $S$ parameter corresponds
to the number of Gaussian standard deviations (reproduced from \cite{bonamente2017book}).}
\label{tab:K-Poisson}
\begin{tabular}{lllllll}
\hline
 & \multicolumn{3}{c}{Lower limit} & \multicolumn{3}{c}{Upper Limit} \\
\hline
Counts   & \multicolumn{3}{c}{Enclosed Probability} & \multicolumn{3}{c}{Enclosed Probability} \\
$n$ & 0.841 & 0.977 & 0.999 &  0.841 & 0.977 & 0.999\\
\hline
 & \multicolumn{3}{c}{Poisson $S$ parameter}  &  \multicolumn{3}{c}{Poisson $S$ parameter} \\
 & 1 & 2 & 3 & 1 & 2 &3 \\
\hline
1 & 0.17 & 0.01 & 0.0 & 3.32 & 5.40 & 7.97 \\
2 & 0.71 & 0.21 & 0.03 & 4.66 & 7.07 & 9.97 \\ 
3 & 1.37 & 0.58 & 0.17 & 5.94 & 8.62 & 11.81 \\
4 & 2.09 & 1.04 & 0.42 & 7.18 & 10.11 & 13.54 \\
5 & 2.85 & 1.57 & 0.75 & 8.40 & 11.55 & 15.19 \\
10 & 6.90 & 4.71 & 3.04 & 14.28 & 18.31 & 22.84 \\
20 & 15.57 & 12.04 & 9.16 & 25.56 & 30.86 & 36.67 \\
\hline
\end{tabular}
\end{table}

For convenience, this note uses the number of equivalent standard deviations of a 
 Gaussian distribution  to describe the probability and confidence intervals of 
$\Delta K$, e.g., a confidence interval with probability 68\% corresponds
to a $\pm 1 \sigma$ interval. 

\subsection{Choice of wavelength range for the search of fluctuations
and the effects of redshift}
\label{sec:N}
In general a spectrum 
is between two wavelengths $\lambda_1$ and $\lambda_2$ and with 
of resolution $\sigma_{\lambda}$.
The total number of independent resolution elements $N$
in the spectrum can be therefore calculated via

\begin{equation}
 N = \frac{\Delta \lambda}{\sigma_{\lambda}},
\label{eq:N}
\end{equation}

where $\Delta \lambda = \lambda_2 - \lambda_1$ is the wavelength range covered
by the observation. Both quantities in the equation above can
be expressed in units of Angstrom (\AA) or, equivalently, meters.
The number $N$ represents the number of opportunities for one fluctuation
to appear in the spectrum, given the instrumental resolution.
For example, a spectrum of $\Delta \lambda =$1~\AA\ range
 with resolution $\sigma_{\lambda}=50$~m\AA\
results in $N=20$ opportunities
for a fluctuation to present itself. This is true only if
there is no \emph{a priori} reason to expect a fluctuation
in a smaller sub--set of the wavelength range covered by the spectrum.

Since fluctuations of interest to this paper are unresolved emission or absorption lines,
it is necessary to examine in more detail at what wavelength(s) 
such lines can be detected.
In general, the wavelength of a given spectral line is known with great accuracy,
as determined by the appropriate (and often quite complicated) quantum mechanical calculation
\citep[e.g.,][]{gatuzz2013}.
For example, the \ovii\ K$\alpha$ line is found at $\lambda_0 = 21.60$~\AA.
Therefore, if one is interested in the detection of 
an unresolved fluctuation corresponding to this line,
there is just $N=1$  opportunity to detect this fluctuation,
and that is given by the
resolution element that includes the $\lambda_0 = 21.60$~\AA\ wavelength.

This situation is complicated by the fact that astronomical lines can
be shifted from their theoretical rest wavelengths
 by Doppler (i.e., velocity) effects or
by the expansion of the universe, the latter causing
a relative velocity between source and observer similar to 
the usual Doppler effect. In both cases, the observer
has a different velocity relative to the source, and this difference
in velocity results in 
a line of rest wavelength $\lambda_0$ being now detected at a new wavelength of

\[ \lambda_1 = \lambda_0 \left( 1 + z\right) \]

where $z=v/c$ is the so--called \emph{redshift} of the source, and $v$ is the
associated relative velocity between source and observer, when $v \ll c$. The reason for the
term \emph{redshift} is that the expansion of the universe causes any
distant source to move away from Earth with a positive velocity ($v > 0$ and $z > 0$),
leading to a shift of visible radiation towards longer wavelengths ($\lambda_1 > \lambda_0$), i.e., towards
the red portion of the spectrum. Incidentally, the redshift is also a measure of the
distance of a source, since the expansion of the universe causes
\emph{farther} sources to move \emph{faster}, i.e., with higher $z$, away from Earth,
according to Hubble's law.

The redshift of the source can have the following implications
for the detection of fluctuations:

(a) If the redshift $z$ is known, then the fluctuation will be expected at a different
wavelength, yet still in  a single resolution element. In this case, it is still
true that $N=1$. For example, this is the case if one has an a priori knowledge of
the location (and therefore redshift) of the astrophysical plasma
responsible for the absorption line. One example of this situation
is the quasar \emph{X Comae}, studied in \cite{bonamente2016}.
The quasar has a known source in its foreground, the \emph{Coma} cluster of galaxies
at $z=0.0231$. Therefore, an absorption line from \ovii\ will now occur exactly 
at $\lambda_1=\lambda_0 (1+0.0231)=22.10$~\AA, instead of 
its rest wavelength of $\lambda_0=21.60$~\AA. There is no 
wavelength uncertainty here
in the search for such absorption line.

(b) When there is no exact knowledge of the location of the source of emission or absorption,
the redshift dependence of the observed wavelength widens the possible range
of wavelengths for the search of fluctuations.
A typical example is that of quasars used to study absorption lines \emph{between}
the redshift of the source ($z_S$) and the observer (by definition at $z=0$). In this
case, the interval of redshifts between 0 and $z_S$ ($\Delta z = z_S$) 
corresponds to a continuous wavelength
range from $\lambda_0$ to $\lambda_0 ( 1 + \Delta z)$, i.e., a wavelength
range of size $\lambda_0 \Delta z$. In this situation, the number of independent
opportunities to detect a feature is

\begin{equation}
 N = \lambda_0 \frac{\Delta z}{\sigma_{\lambda}}.
\label{eq:N-Deltaz}
\end{equation}

In a more general situation, one may have a prior knowledge  that an absorber may
be between redshifts $z_a\geq 0 $ and $z_b \leq z_S$, resulting in the use of
$\Delta z = z_b - z_a > 0$ as the range of redshifts allowed in the search.
For example, this situation was applicable to the study of an extragalactic
source in \cite{fang2002}. A given line can now appear in several resolution
elements, simple because of uncertainties in its location and therefore redshift.

In summary, although the rest wavelength of a line is known with accuracy,
the  \emph{observed} wavelength may have a continuous distribution because of
its redshift dependence. The priors on the location of possible
lines therefore determine the range of wavelengths to use in the search,
and therefore the number of opportunities $N$, according to Eq.~\ref{eq:N-Deltaz}.

\section{Methodology II: Statistical model for the probability of \emph{at least} one fluctuation}
\label{sec:model1}

This section describes a general model to determine the probability
of occurrence of at least one fluctuation. As is usually the case in probability and
statistics, it is necessary to first state as accurately as possible 
the question of interest.
The question initially posed is simply:

\begin{question}
What is the probability of occurrence of 
one chance fluctuation with significance $\geq n\sigma$?
\end{question}

The phrase `\emph{significance $\geq n\sigma$}' is meant as
the probability associated with exceeding  a value of $\pm n$ standard deviations of
a Gaussian distribution.
For example, a $\geq 3\sigma$ fluctuation corresponds to a null 
hypothesis probability of $p=0.003$ or 0.3\% (Table~\ref{tab:K-Gaussian}).
The model is presented and
discussed in Section~\ref{sec:model1-1}, then methods of application are presented
in Section~\ref{sec:model1-2}.

\subsection{Development of the model}
\label{sec:model1-1}
The  model for the probability of an unresolved fluctuation has three components:

(1) First step 
is the evaluation of the
number of available tries, $N$, as discussed in 
Section~\ref{sec:N}. The equations presented in that section can be used
to evaluate the number $N$, according to the situation at hand.

(2) Second, one needs to establish
a level of significance for the fluctuation.
A fluctuation of significance $\geq n\sigma$, say $n=3$,
 means that there is a probability $p$, in this example $p=0.003$ or 0.3\%, 
of occurrence of one such fluctuation
at this this level of probability. 
This step assumes that there is interest
in both positive and negative fluctuations, i.e., the null 
hypothesis is two--sided. 
If $\Delta K$ is not Gaussian, the appropriate distribution
must be used to determine the
probability of interest $p$, as described in Section~\ref{sec:DeltaK}.

\begin{figure}[!t]
\centering
\includegraphics[width=2.8in]{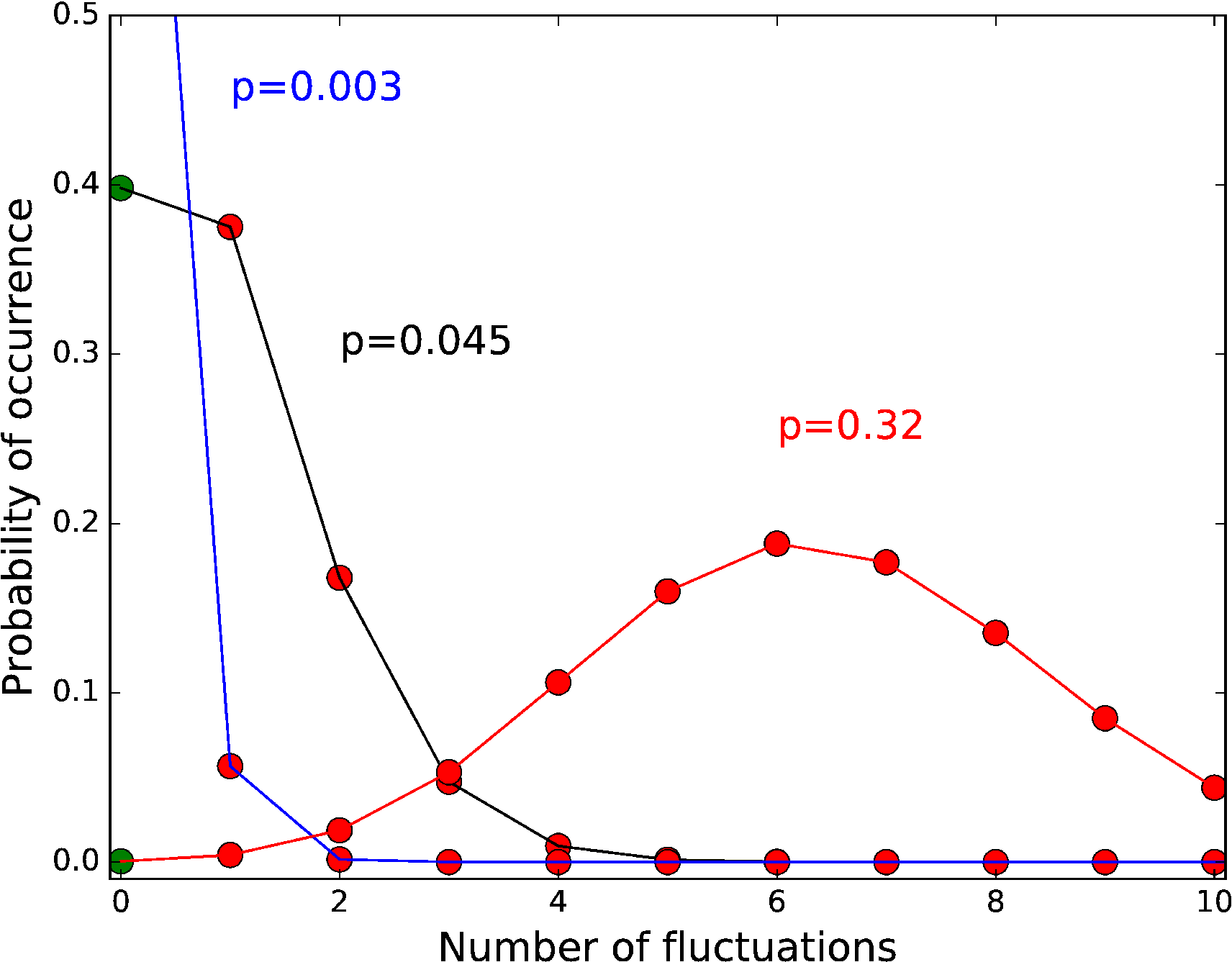}
\includegraphics[width=2.8in]{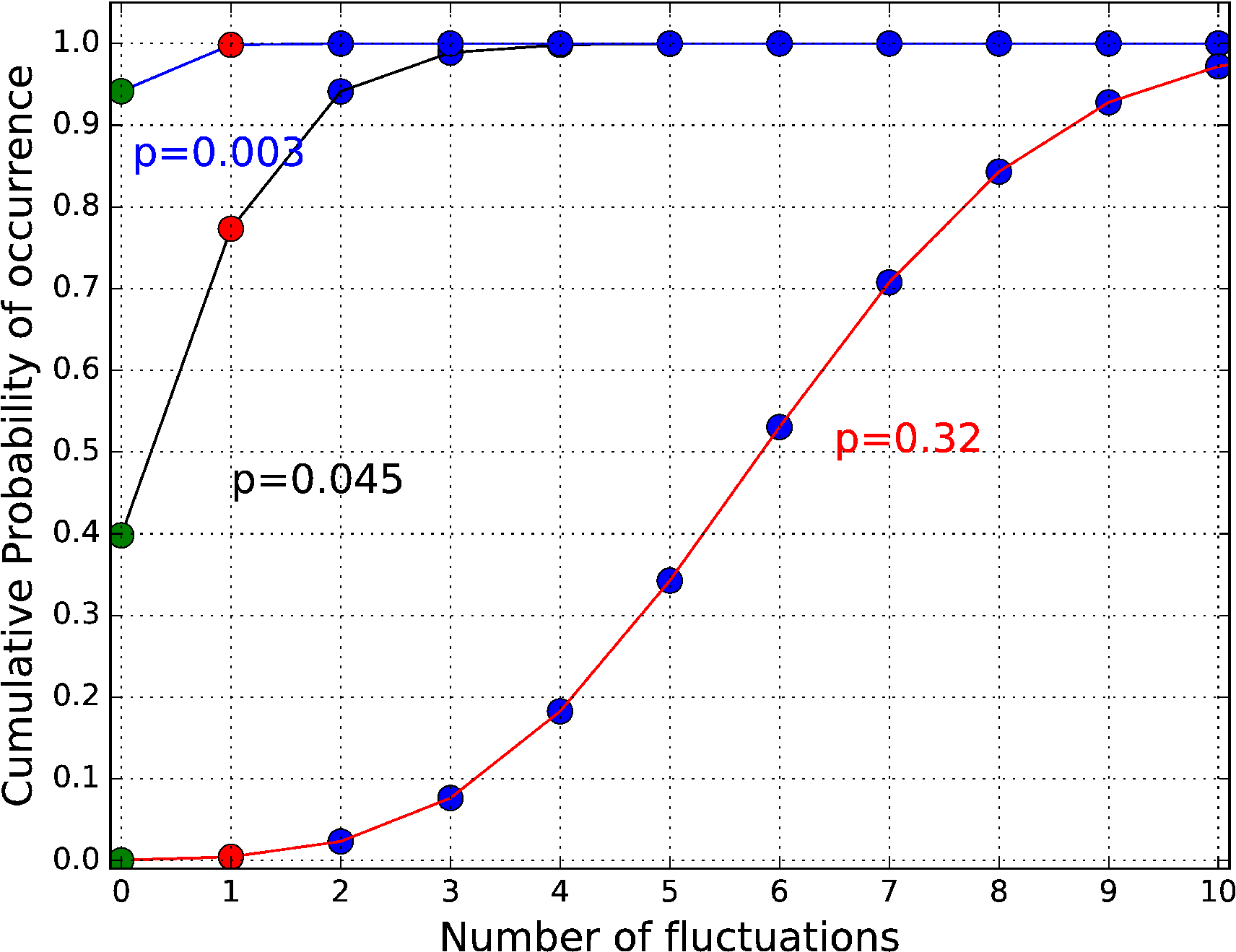}
\caption{\small (Left) Probability of occurrence of number of fluctuations, 
for three representative values of probability $p$ 
 and fixed $N=20$. The sum of probabilities with red dots 
is the cumulative probability of at least one such fluctuation.
(Right) Cumulative distribution functions for the same distributions as in the left panel.
Each dot represents the probability of \emph{at least} that number of fluctuations, or larger.}
\label{fig:pi}
\end{figure}

(3) The binomial distribution is used to describe the probability of occurrence of
one such fluctuation out of $N$ possible tries, via
\begin{equation}
P(1) = \binom{N}{1} p q^{N-1} = N p q^{N-1}
\label{eq:p1}
\end{equation}
where $p$ is the null hypothesis probability for the occurrence of a
fluctuation, and $q=1-p$ is the probability of the complementary event, i.e., the non--occurrence of such 
fluctuation. For example, according to Equation~\ref{eq:p1} and for $N=20$, 
one $\geq 1\sigma$ fluctuation ($p=0.32$) 
has just a 0.4\% probability of occurring one time,
and one $\geq 2\sigma$ fluctuation ($p=0.045$) 
has a 38\% probability of occurring one time, 
according to Equation~\ref{eq:p1}.
At first, this example seems counter--intuitive, as one 
might expect that a lower--significance fluctuation would be more likely than
a higher--significance fluctuation.
However, Eq.~\ref{eq:p1}
only calculates the occurrence of \emph{exactly} one such fluctuation. Surely, one expects more than one 
$\geq 1\sigma$ fluctuations in a spectrum, especially over a wide wavelength coverage,
as shown in Figure~\ref{fig:pi}. 

This observation leads to a reformulation of the original question to include the 
statement of \emph{at least one} instead of \emph{exactly one} fluctuation.
Therefore the statistically interesting question becomes:

\begin{question}
What is the probability of occurrence 
of at least one chance fluctuation with significance $\geq n\sigma$?
\end{question}

This probability is given by
\begin{equation}
P = \sum_{i=1}^N P(i),
\label{eq:p}
\end{equation}
where 
\begin{equation}
P(i) = \binom{N}{i} p^i q^{N-i}.
\label{eq:pi}
\end{equation}
The cumulative probability $P$ of having at least one fluctuation
is now a monotonic function of the individual probability $p$, which is in
turn a monotonic function of the number $n$ of Gaussian standard deviations
to be exceeded by the fluctuation.  In other words, Equation~\ref{eq:p}
ensures a meaningful null hypothesis probability, i.e., the null
hypothesis probability $P$ of \emph{at least} one fluctuation
is monotonically decreasing as $p$ decreases (for same $N$, see 
Table~\ref{tab:PpN}). 
{\blue
According the Equation~\ref{eq:p} (which replaces Equation~\ref{eq:p1})
and Table~\ref{tab:PpN}, for $N=20$,
at least one $\geq 1 \sigma$ fluctuation has now a $\sim 99.9\%$
(versus 0.4\% of exactly one fluctuation), and at least one $\geq 2 \sigma$
fluctuation  has a $\sim 60\%$ probability (versus 38\% of exactly one fluctuation).
}
An illustration of the binomial probabilities of Equation~\ref{eq:pi} is shown in Figure~\ref{fig:pi}.

\begin{figure}[!t]
\centering
\includegraphics[width=2.8in]{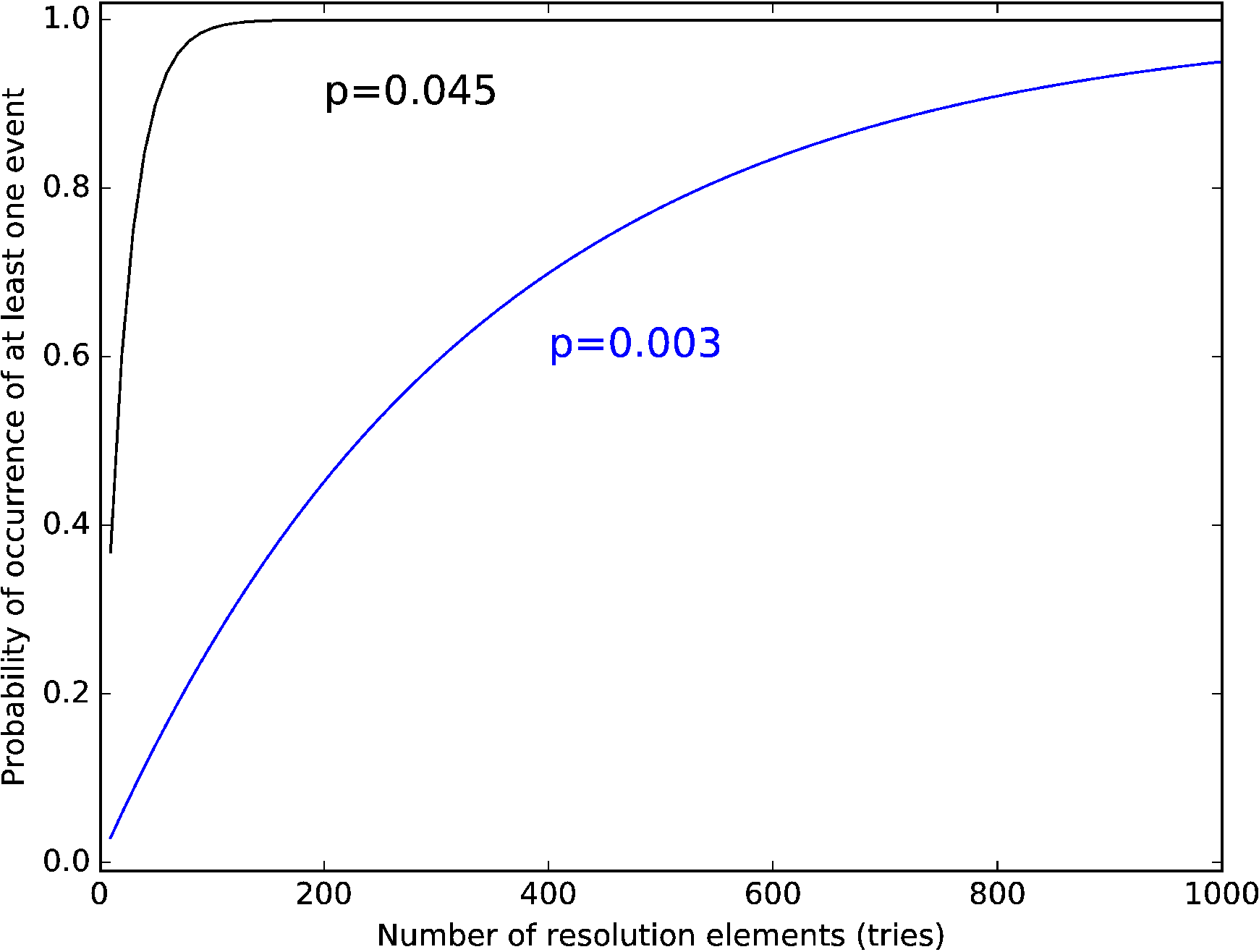}
\includegraphics[width=2.8in]{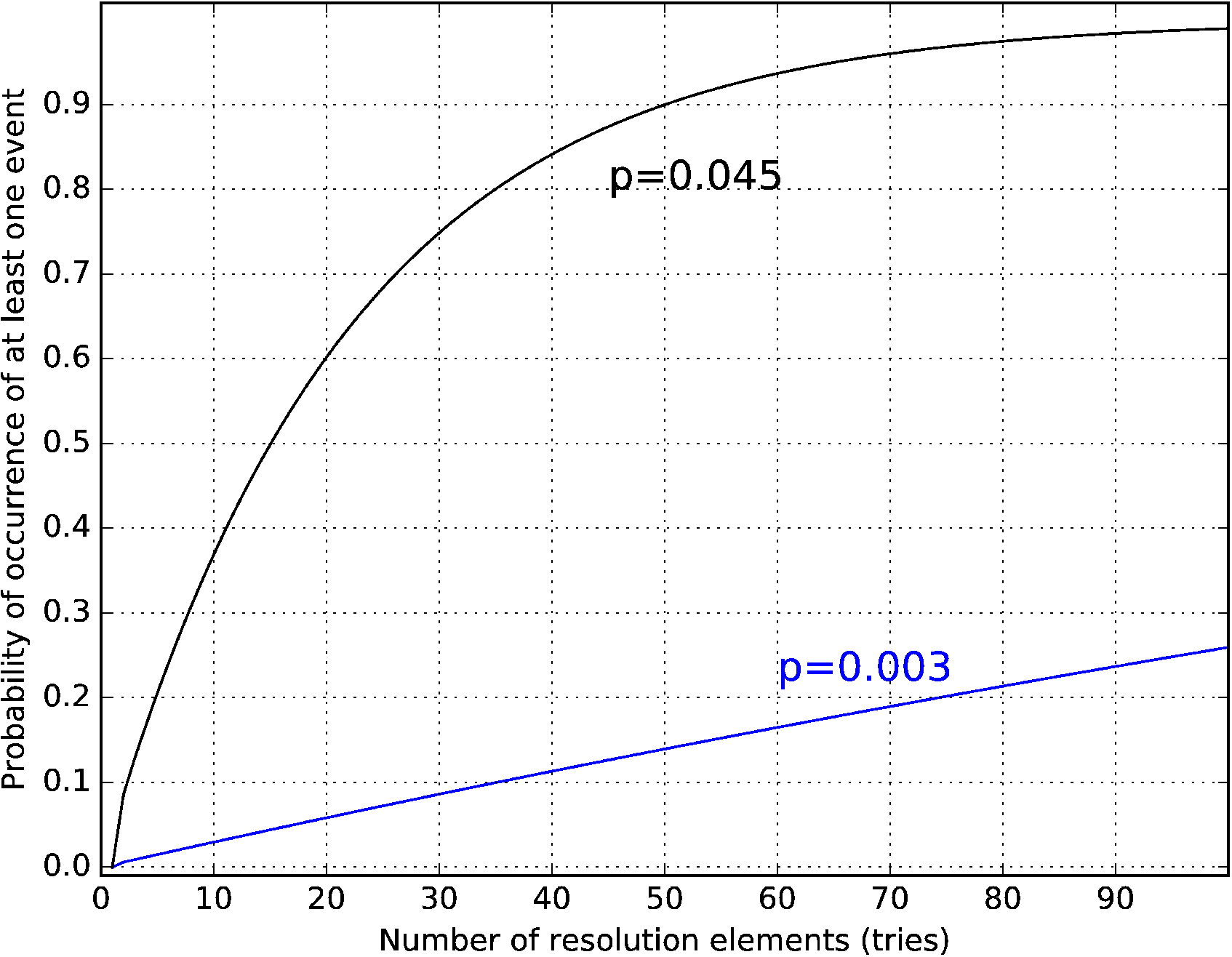}
\caption{\small (Left:) Cumulative probability of occurrence 
of at least one occurrence, 
for two representative values of the confidence
level. (Right:) Zoom--in to small values of the number of resolution elements. }
\label{fig:P}
\end{figure}

The cumulative probability $P$ of at least one chance fluctuation
 is a function of both $N$ and $p$, and it is shown for
two representative values of $p$ in Figure~\ref{fig:P} as function of $N$. 
For example, there is approximately 
a 6\% probability of having at least one $\geq 3 \sigma$ fluctuation
in $N=20$ resolutions elements.
\begin{table}[t]
\centering
\caption{Cumulative probabilities P of having at least one fluctuation as function of 
probability $p$ and number of 
resolution elements $N$.}
\footnotesize
\begin{tabular}{ll|lllllllll}
\hline
\multicolumn{2}{c}{Confidence} & $N=5$ & $N=10$ & $N=20$ & $N=40$ & $N=60$ & $N=100$ & $N=300$ & $N=1000$\\
$p$ & Equiv. $\sigma$ & \\
\hline
0.32 & $\pm 1 \sigma$	& 0.8546 & 0.9789 & 0.9996 & 1.0    & 1.0    & 1.0      & 1.0 & 1.0 \\
0.1  & $\pm 1.6 \sigma$	& 0.4095 & 0.6513 & 0.8784 & 0.9852 & 0.9982 & 1.0	& 1.0 & 1.0\\
0.045 & $\pm 2 \sigma$ 	& 0.2056 & 0.3690 & 0.6018 & 0.8415 & 0.9369 & 0.9900	& 1.0 & 1.0 \\
0.01  &$\pm 2.6 \sigma$	& 0.0490 & 0.0956 & 0.1821 & 0.3310 & 0.4528 & 0.6340	& 0.9510 & 1.0 \\
0.003 &$\pm 3 \sigma$	& 0.0149 & 0.0296 & 0.0583 & 0.1132 & 0.1650 & 0.2595 & 0.5940 	& 0.9504\\
0.001  &$\pm 3.2 \sigma$& 0.0050 & 0.0100 & 0.0198 & 0.0392 & 0.0583 & 0.0952	& 0.2593 & 0.6323\\
0.0001 &$\pm 3.9 \sigma$& 0.0005 & 0.0010 & 0.0020 & 0.0040 & 0.0060 & 0.0100  & 0.0296 & 0.0952 \\
0.000064 & $\pm 4 \sigma$ & 0.0003 & 0.0006 & 0.0013 & 0.0025 & 0.0038 & 0.0063 & 0.0187 & 0.0611 \\
0.000006 & $\pm 4.5 \sigma$ & 0.0000 & 0.0001 & 0.0001 & 0.0002 & 0.0004 & 0.0006 & 0.0018 & 0.0060 \\
\hline
\end{tabular}
\label{tab:PpN}
\end{table}
Probabilities of having at least one fluctuation are also summarized in Table~\ref{tab:PpN}.
This table will guide the data analyst 
to choose the proper significance level $p$ of unresolved features
of interest. For a search with $N=100$ resolution elements, one expects at least one 
fluctuation that exceeds the $2 \sigma$ level with 99\% probability. This is to say, there
is a near certainty that there will be at least one fluctuation at this level simply
by chance, according to the uncertainties in the measurements.
At the $3 \sigma$ level, this probability is still approximately 26\%,
{\blue and at the 4 $\sigma$ level it is reduced to just 0.6\%. 
In a blind search for fluctuations
with $N=100$ resolution elements, an analyst would therefore need to identify 
a fluctuation with at least (approximately) 4 $\sigma$ significance,
according to Table~\ref{tab:PpN}.}

\subsection{Methods of application of the model}
\label{sec:model1-2}

Consider finding one $\geq 3 \sigma$ fluctuation
in a spectrum with 100 resolution elements  
during a \emph{blind} search in which the feature can equally well appear in 
any of the resolution elements.
Such feature should be assigned a chance probability of 26\%. Alternatively, one can
say that there is a 26\% probability that such feature is \emph{not} real and it is
caused by randomness in the measurements. This effectively means that even a $3 \sigma$ 
fluctuation cannot be trusted with great confidence, if it is the result of a 
search with such large number of opportunities for detection.
A careful analyst will \emph{not} claim this detection as real.
With 100 resolution elements available, an individual $\geq 4\sigma$ detection results in
a cumulative null hypothesis probability of $p=0.0063$, i.e., there is less than a 1\%
chance of such fluctuation being caused by randomness in the measurements.
It would appear reasonable to claim such a detection as real.

In general there are two types of application of the model for one unresolved
fluctuation:

{\blue (a) The search for a line of \emph{known wavelength} and in a fixed redshift range. 
}

This is the case, for example,
of the search for a \ovii\ K$\alpha$  absorption line, with rest wavelength of
$\lambda_0$=21.60~\AA. 
The relevant equation for $N$ is Equation~\ref{eq:N-Deltaz}, with
$z_a$ and $z_b$ corresponding to the range of redshift where the line is expected.
Notice that even when one expects a line at a specific location, e.g., due by 
a line--of--sight cluster of known redshift, one should still include a
redshift range to allow for errors in the measurements of the redshift.
This redshift error may be small, and in that case the equation will yield $N\leq 1$,
which should be interpreted as $N=1$.

{\blue (b) The search for a new line of \emph{unknown wavelength}. 
}

This is the case when
a fluctuation appears at a wavelength that is not consistent with any known
lines. In this case, the burden in the calculation of the appropriate value of $N$
is that of excluding all redshift ranges where all known lines may appear.
Having established the range $\Delta \lambda$ where no known lines are present
(or a union of several such ranges), the number of independent tries for the detection
of an unknown line will be given by Equation~\ref{eq:N}.
For example, there has recently been a claim of detection of a new emission
line due to the presence of dark matter during a blind search
of X--ray spectra of galaxy clusters \citep{bulbul2014}. 
In this work the authors calculate the number of
resolution elements available in the search, correctly excluding a portion of the
spectrum 
where several known lines are known to be present.

\section{Methodology III: Statistical model for the probability of \emph{at least} more than one fluctuation}
\label{sec:model2}
Another situation of practical interest is the search for \emph{more than one} 
unresolved features in the same spectrum. This section extends the model
presented in Section~\ref{sec:model1} to the case of several fluctuations.

\subsection{Model for more than one fluctuation}
\label{sec:model2-1}
In this case, the statistical question
of interest becomes:

\begin{table}[!t]
\centering
\caption{Cumulative probability P of having at least \emph{two} fluctuations as function of
confidence (or probability) level $p$, and number of
resolution elements (or tries) $N$.}
\footnotesize
\begin{tabular}{l|llllllllll}
\hline
Confidence & $N=2$ & $N=5$ & $N=10$ & $N=20$ & $N=40$ & $N=60$ & $N=100$ & $N=300$ & $N=1000$\\
$p$ & \\
\hline
0.32   & 0.1024	& 0.5125 & 0.8794 & 0.9953 & 1.0000 & 1.0000 & 1.0000 & 1.0000 & 1.0000 \\
0.1    & 0.0100 & 0.0815 & 0.2639 & 0.6083 & 0.9195 & 0.9862 & 0.9997 & 1.0000 & 1.0000 \\
0.045  & 0.0020 & 0.0185 & 0.0717 & 0.2266 & 0.5426 & 0.7584 & 0.9428 & 1.0000 & 1.0000\\
0.01   & 0.0001 & 0.0010 & 0.0043 & 0.0169 & 0.0607 & 0.1212 & 0.2642 & 0.8024 & 0.9995 \\
0.003  & 0.0000 & 0.0001 & 0.0004 & 0.0016 & 0.0065 & 0.0142 & 0.0367 & 0.2275 & 0.8013\\
0.001  & 0.0000 & 0.0000 & 0.0000 & 0.0002 & 0.0008 & 0.0017 & 0.0046 & 0.0369 & 0.2642\\
0.0001 & 0.0000 & 0.0000 & 0.0000 & 0.0000 & 0.0000 & 0.0000 & 0.0000 & 0.0004 & 0.0047 \\
\hline
\end{tabular}
\label{tab:PpN-2}
\end{table}

\begin{table}[!t]
\centering
\caption{Cumulative probability P of having at least \emph{three} fluctuations as function of
confidence (or probability) level $p$, and number of
resolution elements (or tries) $N$.}
\footnotesize
\begin{tabular}{l|lllllllll}
\hline
Confidence & $N=3$ & $N=5$ & $N=10$ & $N=20$ & $N=40$ & $N=60$ & $N=100$ & $N=300$ & $N=1000$\\
$p$ & \\
\hline
0.32	& 0.0328 & 0.1905 & 0.6687 & 0.9765 & 1.0000 & 1.0000 & 1.0000 & 1.0000 & 1.0000 \\
0.1     & 0.0010 & 0.0086 & 0.0702 & 0.3231 & 0.7772 & 0.9470 & 0.9981 & 1.0000 & 1.0000 \\
0.045   & 0.0001 & 0.0009 & 0.0086 & 0.0586 & 0.2681 & 0.5103 & 0.8328 & 0.9999 & 1.0000\\
0.01    & 0.0000 & 0.0000 & 0.0001 & 0.0010 & 0.0075 & 0.0224 & 0.0794 & 0.5779 & 0.9973 \\
0.003   & 0.0000 & 0.0000 & 0.0000 & 0.0000 & 0.0002 & 0.0008 & 0.0035 & 0.0626 & 0.5771\\
0.001   & 0.0000 & 0.0000 & 0.0000 & 0.0000 & 0.0000& 0.0000 & 0.0002 & 0.0036 & 0.0802\\
\hline
\end{tabular}
\label{tab:PpN-3}
\end{table}

\begin{question}
What is the probability of occurrence of at least $m \geq 2$ fluctuations 
with significance $\geq n\sigma$?
\end{question}

The number of  trials $N$ is calculated following the same steps
as in Section~\ref{sec:model1}.
The question can be answered by a straightforward
modification of Equation~\ref{eq:p}:
\begin{equation}
P_m = \sum_{i=m}^N P(i),
\label{eq:pm}
\end{equation}
to account for the fact that $m\geq 2$ fluctuations are the new null hypothesis.
The model for more than one fluctuation is therefore a simple extension of the
model for just one fluctuation. Tables of probabilities for the cases of $m=2$ and
$m=3$ are reported respectively in Tables~\ref{tab:PpN-2} and \ref{tab:PpN-3}.
For example, the probability of at least
$m=3$ random fluctuations at $\geq 2 \sigma$ for $N=10$ is 0.86\% (Table~\ref{tab:PpN-3}),
it is 7.2\% for at least $m=2$ fluctuations  (Table~\ref{tab:PpN-2}) 
and 36.9\% for at least one fluctuation (Table~\ref{tab:PpN}).

It is important to notice that this model considers $m$ fluctuations with the
\emph{same} minimum statistical significance $n \sigma$. This is typically
a reasonable assumption, when performing a blind search for lines of unknown intensity.

\subsection{Methods of application of the $m\geq2$ model}
\label{sec:method2-app}
There are three main applications of this model:

(a) The search for $m\geq2$ lines of {\blue fixed relative distance}.

Consider the case of a plasma with fixed thermodynamical properties, such
as fixed temperature and density. Radiative transfer codes can predict
the occurrence of several absorption lines at specified wavelengths, optionally also
with specified line intensity ratios. For example, a hot plasma 
may have \emph{both} \ovii\ and 
\oviii\ K$\alpha$ lines, respectively at 21.60 and 18.97~\AA\ in the rest frame,
with a specified ratio of intensity that depends on temperature.
Another example is that of Lyman--$\alpha$ and Lyman--$\beta$ lines
from the same ion, which occur at prescribed wavelengths and intensity for
all ions.
The relative location and intensity of the lines is fixed and therefore
they should be considered as part of a \emph{single} spectral feature, 
spread over the appropriate wavelength range. 
In this case, the relevant model becomes that of Section~\ref{sec:model1}
for just one fluctuation. 

The problem  therefore shifts to determining the desired \emph{combined} significance
$p$ for the two (or more) lines. For example, consider two lines 
of identical intensity detected
at a significance of $3 \sigma$ each. If the lines are detected independently,
then the combined significance of two 3--$\sigma$ lines is obtained by quadrature
addition, leading to a 4.2 $\sigma$ significance (this is true only if fluctuations
have the same intensity or counts). This is equivalent to a re-binning
of the data such that the two features would appear in the same bin (i.e.,
counts in the two bins are added). The probability of chance detection
of two such lines can be therefore calculated using the $p$ value that corresponds to 
the combined significance of 4.2~$\sigma$.

The next step is the calculation of the number of trials $N$. 
Same considerations as discussed in Section~\ref{sec:model1} apply.
If lines
are at a fixed wavelength, and fixed redshift, then typically $N=1$.
In the example above of two 3--$\sigma$ lines, the chance probability
is just that of a 4.2 $\sigma$ fluctuation, which is less than 0.01\%.
If there is a range of redshifts to be considered, then one would 
use Equation~\ref{eq:N-Deltaz} with $\Delta z$ corresponding 
to a range of wavelengths that covers all lines under consideration,
and use the number $N$ of tries in the calculation of the probability.

This case  applies
when fitting a spectrum to a model for several lines, for example,
as done by \cite{nevalainen2017a} to model absorption lines
from various elements at the same temperature.
 Since all
lines are related, typically the only free parameter in the fit is the
intensity of one of the lines (or an equivalent parameter), 
with other lines having intensities linked to 
that of the reference line. All lines contribute to the statistical
significance of detection of this one--parameter model. 

(b) The search for $m\geq2$ lines {\blue at unknown relative distance}.

This case applies to the search of lines with
known wavelengths in the rest frame, but whose redshifts are unknown and 
thus the lines
don't have fixed distance from one another, as in case (a). 
This case is applicable to the blind  search for, e.g., several
\ovii\ K$\alpha$ lines without having priors on redshifts, as recently
done in a paper claiming the detection of the
missing baryons via X--ray spectroscopy \cite{nicastro2018}.  We illustrate
the application of this method to three typical situations, (1) one in which $m=2$
absorption lines have been detected, (2) with
an analysis on the results of \cite{nicastro2018}, and (3) an example for $m=3$. 

(1) For $m=2$ and $N=100$ tries, there is a null hypothesis probability of
$P=0.037$ or 3.7\% that two $\geq 3 \sigma$ fluctuations occur, as given
by 
Table~\ref{tab:PpN-2}. One would conclude that it is possible that these two
detected fluctuations are random, and neither should be claimed as real.
But if there were only $N=20$ tries, or resolution elements, then the probability
would drop to 0.16\%. One would conclude that the two fluctuations are unlikely
due to randomness, and \emph{at least} one is real. If only one fluctuation
had been detected instead of two, Table~\ref{tab:PpN} indicates that for $N=20$
tries there is a 5.8\% null hypothesis probability that the signal is due to
a random fluctuation. So, that fluctuation would not be claimed as real. 
Therefore, as expected, the simultaneous detection of two fluctuations is a 
much more statistically significant result than a single fluctuation, 
and the models presented in this paper
permit a quantitative comparison between the two cases.

(2) In \cite{nicastro2018} is reported the detection of two redshifted \ovii\ lines 
in the line of sight towards a distant X--ray bright source. The importance of this
result is that it indicates the discovery of hitherto \emph{missing} baryonic 
matter, possibly closing a long--standing mystery in modern cosmology \cite{cen1999}.
The finding of this paper is that two \ovii\ K$\alpha$ lines have been identified,
respectively at $z=0.355$ with significance 3.7-4.2 $\sigma$ (the range
indicates uncertainties in the data analysis) and $z=0.434$ with significance
4.1-4.7 $\sigma$. Given the wavelength range of the blind search,
the authors indicated that there were approximately $N \simeq 400$ resolution elements,
having excluded regions of known artifacts or unrelated lines from the search.
The authors conducted their own analysis
to account for the redshift trials in the blind search, 
based on Monte Carlo simulations, and concluded that these
lines remain significant even after accounting for the number of redshift trials.
These lines lend themselves to an analysis of their significance using the methods
presented in this paper. For $m=2$ (two lines detected) with significance 
$\geq 3.9 \sigma$ (using a reasonable approximation of the results in \cite{nicastro2018}),
or $p=1\times 10^{-4}$, Table~\ref{tab:PpN-2} indicates
a cumulative null hypothesis probability of $\sim$0.04\%. Clearly these two detections
cannot be given to random fluctuations, even after accounting for the
number of available resolution elements, thus confirming the conclusion obtained
by the authors based on their Monte Carlo simulations. At least one of these two
lines must be a real astrophysics signal.
Had only one line (instead of two) been detected with significance $\geq 3.9 \sigma$,
then Table~\ref{tab:PpN} for $N=300$ would indicate a null hypothesis probability of
3\%. In this case, one would be much less confident to claim the detection of that
lone line.

(3) For $m=3$ and $N=100$ tries, there is a large probability of at least
three $\geq 2 \sigma$ fluctuations (83.3\%). An analyst finding three such fluctuations
in a blind search should therefore assume that these are chance fluctuations.
Not so for $\geq 3 \sigma$ fluctuations, since there is now just a 0.35\% 
probability of these simultaneous occurrences being caused by chance fluctuations.
In this case, it is still possible that perhaps one of the three fluctuations is caused
by random effects. Yet, the simultaneous occurrence of three fluctuations is unlikely
to be caused by randomness, so that at least one or two of those
fluctuations are likely to be real.
The same arguments can be repeated for any value of $m$, 
and tables equivalent to Table~\ref{tab:PpN-3}
can be constructed from Equation~\ref{eq:pm}.

(c) The search for $m \geq 2$ lines of both known and unknown relative 
distance.

This is probably the most common occurrence in X--ray astronomy. Typical
X--ray sources have 
a handful of $z=0$ absorption lines (originating from the Galaxy) that are often detected
with high significance, on occasion accompanied by few faint absorption--line features
that occur at random wavelengths. This was the case for the source in \cite{nicastro2018}
presented above
in case (b), where the 
two $z>0$ absorption lines were accompanied by several 
stronger absorption lines from Galactic material (such as \ovii, \oi, and many more ions).
In such cases, it is customary to first identify and model the 
lines at fixed wavelength, and then proceed with the blind search of additional lines
as described in case (b). The modeling of bright lines at expected wavelengths, such
as \ovii\ at 21.6~\AA, is also needed to provide an accurate model
for the continuum, prior to the search for fainter lines. When the blind search commences,
the wavelength range to be searched, and the associated number of resolution elements,
is reduced to exclude the range where the fixed--wavelength lines were previously identified.
This procedure was in fact followed by \cite{nicastro2018} and for
the source in the case study of Section~\ref{sec:pg1116} \cite{bonamente2016}.
In so doing, the fluctuations corresponding to expected features are incorporated in
the search for new fatures.

\section{Methodology IV: The approximate $p N$ method}
\label{sec:approximation}
A common and simplified method to deal with the number of resolution elements 
in blind searches is that
of \cite{nicastro2013}, also implemented by \cite{bonamente2016} and others.
For the detection of one feature, the null hypothesis probability
$p$ of detection (e.g., $p=0.003$ for a $3\sigma$
detection) is simply multiplied by the number of resolution elements $N$,

\[ P = p N, \]

to obtain the cumulative probability $P$ of the null hypothesis.
This empirical equation is intended to provide
 the probability of detection of one feature,
adjusted for the number of possible detection opportunities, or
resolution elements. 

This probability can be shown to be
a good approximation to Equation~\ref{eq:p1} when $p \ll 1$ and $N$ is not too large.
The main drawback of this approximation is that
it does not ensure  the property of $P \leq 1$,
which is one of the fundamental axioms of 
the theory of probability. Therefore it
is preferable to use Equation~\ref{eq:p1} to address
the probability of detection of exactly one unresolved feature,
or Equation~\ref{eq:p} 
for at least one unresolved feature. 
Another  limitation is that this simple method does not calculate the
probability of \emph{at least} one fluctuation, but that of 
\emph{exactly} one fluctuation.

It is useful to investigate when this simplified $pN$ method can be used
in place of Equation~\ref{eq:p}, to approximate the
probability of detection of at least one fluctuation with
individual significance $p$, out of $N$ tries.
For this purpose, the probability of having at least one
fluctuation of significance $p$ can be expanded as a sum,

\[ P(\geq 1) = N p q^{N-1} + \frac{N (N-1)}{2} p^2 q^{N-2} 
+ \dots + \binom{N}{i}  p^i q^{N-i} + \dots \; (i \leq N). \]

For the approximation to hold, one needs to ensure that
there is a much smaller
probability of two fluctuations than just one
fluctuation. This means that:

\[ \frac{N (N-1)}{2} p^2 q^{N-2} \ll N p q^{N-1} \]

If $p$ is small, $p \ll 1$,  this condition becomes

\begin{equation}
p N \ll 2.
\label{eq:appr1}
\end{equation}
The condition of $p \ll 1$ is quite reasonable, since the analyst usually seeks
fluctuations with a small null hypothesis probability in the first place.
The resulting condition given by Equation~\ref{eq:appr1} is somewhat more constraining.
This condition can be rephrased by saying that, for a given confidence level $p$,
the number of tries must be \emph{small enough} to satisfy the inequality;
alternatively, for a give number of tries $n$, the individual
level of probability $p$ must  be \emph{small enough} 
to satisfy the condition.
For example, in Table~\ref{tab:PpN}, all the entries for the first line
do \emph{not} satisfy this criterion, while the entries in the
last line \emph{do} satisfy the criterion.

Next, one also needs to ensure that the probability of more than two fluctuations
are likewise negligible, compared to the probability of one fluctuation. This  condition
can be enforced by requiring that

\[ P(i+1) \ll P(i)  \;  ,\text{for }i \geq 2\] 

Using the binomial distribution for the probability $P(i)$, 
it is easy to show that this condition becomes

\[ p N \ll (i+1). \]

This condition for $i \geq 2$ is automatically satisfied if Equation~\ref{eq:appr1} applies,
therefore the only conditions for the applicability of the approximate
$pN$ method are: 

\[ p \ll 1 \;\; \text{and} \;\; p N \ll 2. \]
This means that there are many situations in which this simplified method can be used, in place of the
more general method described by Equation~\ref{eq:p}. In those cases,
the use of the $pN$ method is justified and it leads to a simpler method of analysis.
The accuracy of the $pN$ approximation can be calculated as function of the
values of $p$ and $N$ (see Table~\ref{tab:approx}).

\begin{table}[!ht]
\centering
\caption{Absolute and percent errors in the use of the $pN$ approximation
for the cumulative probabilities $P$ of having at least one fluctuation as function of
 $p$ and 
 $N$. Exact probabilities $P$ are in Table~\ref{tab:PpN}.}
\footnotesize
\begin{tabular}{ll|lllllllll}
\hline
\multicolumn{2}{c}{Confidence} & $N=5$ & $N=10$ & $N=20$ & $N=40$ & $N=60$ & $N=100$ & $N=300$ & $N=1000$\\
$p$ & Equiv. $\sigma$ & \\
\hline
0.003 &$\pm 3 \sigma$   & 0.0001 & 0.0004 & 0.0017 & 0.0068 &  0.0150 & 0.0405 & 0.3060 & 2.0496 \\
	& & (0.60\%) & (1.36\%) & (2.88\%) &  (5.97\%) & (9.12\%) & (15.60\%) & (51.52\%) & (215.64\%) \\
0.001  &$\pm 3.2 \sigma$&  0.0000 & 0.0000 & 0.0002 & 0.0008 & 0.0017 & 0.0048 & 0.0407 & 0.3677 \\
	& &  (0.20\%) & (0.45\%) & (0.95\%) & (1.96\%) & (2.98\%) & (5.03\%) & (15.70\%) &(58.15\%) \\
0.0001 &$\pm 3.9 \sigma$&  0.0000 & 0.0000  & 0.0000 & 0.0000 & 0.0000  & 0.0000 & 0.0004 & 0.0048 \\
	& &  (0.02\%) &  (0.05\%) &  (0.10\%) &  (0.20\%) &  (0.30\%) & (0.50\%) &  (1.50\%) &  (5.08\%) \\
0.000064 & $\pm 4 \sigma$ & 0.0000 & 0.0000 & 0.0000 & 0.0000 & 0.0000	& 0.0000 & 0.0002 & 0.0019  \\
	& & (0.01\%) & (0.03\%) & (0.06\%) & (0.12\%) & (0.19\%)  & (0.31\%) & (0.94\%) & (3.18\%)  \\
0.000006 & $\pm 4.5 \sigma$ & 0.0000 & 0.0000 & 0.0000 & 0.0000 & 0.0000 & 0.0000 & 0.0000 & 0.0000  \\
	& & (0.00\%) & (0.00\%) & (0.01\%) & (0.01\%) & (0.02\%)  & (0.03\%) & (0.09\%) & (0.30\%)  \\
\hline
\end{tabular}
\label{tab:approx}
\end{table}

\section{Application: A case study from the Bonamente et al. (2016) 
absorption lines in the quasar \pg}
\label{sec:pg1116}

The methods of analysis of fluctuations described in 
Sections~\ref{sec:model1}--\ref{sec:approximation} 
are illustrated using the X-ray data of \cite{bonamente2016} for the astronomical
source \pg.

\subsection{Description of the data}
An X--ray  spectrum of the quasar \pg\ was analyzed in search of two 
extragalactic absorption lines, the K$\alpha$ lines from \ovii\ and \oviii,
respectively with a rest wavelength of $\lambda_0$=21.60 and 18.97 \AA.
The search was performed in the neighborhood of 6 different fixed redshifts,
chosen from prior knowledge of
related absorption lines in the far--ultra violet (FUV) portion
of the spectrum \citep{tilton2012}.
This resulted in a total of 12 possible redshift intervals
searched for the presence of absorption lines. The
resolution of the X-ray spectra ($\sigma_{\lambda} \sim 50$~m\AA) are larger than the width
of the lines ($d \lambda \sim 10$~m\AA),
therefore absorption lines would appear as unresolved fluctuations. 

The search was intended to detect
lines originating from a possible
warm--hot intergalactic medium (WHIM) along the sight line towards \pg. 
This WHIM is assumed to be hot enough so that oxygen is ionized
in the form of \ovii\ (six--times ionized)  
and \oviii\ (seven--times ionized) ions. Main results from the \cite{bonamente2016}
paper are reported in Table~\ref{tab:bonamente2016}.
Of the 12 possible lines, \oviii\ at $z=0.1385$ falls at the same wavelength
as \ovii\ at $z=0$ (not shown in table), which may originate in the Galaxy.
Therefore only $N=11$ redshift intervals were useful for the search
of extra--galactic WHIM.

The redshifts of the lines were fixed at the values
indicated by the FUV data, in the initial spectral analysis. In the
case of the 
only $\geq 3 \sigma$ absorption line detected (\oviii\ at $z=0.0928$), the redshift
was subsequently left free to adjust itself to have a more accurate measurement, given the
initial tentative detection. 
The best--fit redshift of \oviii\ at  $z=0.0911$ is consistent with the fixed
value initially used, and  the effect of such redshift adjustment is 
described in Section~\ref{sec:z}. 

\begin{table}[!t]
\small
\centering
\caption{Measurement of features in the \chandra\ spectrum of \pg,
reproduced from \cite{bonamente2016}.}
\label{tab:bonamente2016}
\begin{tabular}{lccc}
\hline
\hline
Line ID & Redshift & Flux $\Delta K$ & Significance\\
 	&	   & ($10^{-6}$ phot cm$^{-2}$~s$^{-1}$) & ($\Delta K/\sigma_K$) \\
\hline
\ovii\ K$\alpha$ & 0.041 & -$7.7\pm7.2$         & -1.1\\
\ovii\ K$\alpha$ & 0.059 & $7.4\pm8.9$          & +0.8\\
\ovii\ K$\alpha$ & $0.0911\pm0.0004$ & $-11.0\pm7.5$ & -1.5\\
\ovii\ K$\alpha$ & 0.1337 & $-3.1\pm8.6$        & -0.4\\
\ovii\ K$\alpha$ & 0.1385 & $39.1\pm11.1$       & +3.6\\
\ovii\ K$\alpha$ & 0.1734 &  $8.4\pm10.1$       & +0.8\\
\hline
\oviii\ K$\alpha$ & 0.041 & $-6.8\pm6.8$        & -1.0\\
\oviii\ K$\alpha$ & 0.059 & $-2.5\pm7.3$        & -0.3\\
\oviii\ K$\alpha$ & $0.0911\pm0.0004$ & $-31.4\pm6.0$& -5.2\\
\oviii\ K$\alpha$ & 0.1337 & $-2.2\pm8.3$       & -0.3\\
\oviii\ K$\alpha$ & 0.1385 & \nodata                &\nodata\\
\oviii\ K$\alpha$ & 0.1734 &$1.4\pm8.9$         & +0.2\\
\hline
\hline
\end{tabular}
\end{table}

\subsection{
Detection of an absorption line from the WHIM towards
\pg}
\label{sec:pg-1}

The first point in the analysis of the spectrum is to
establish the statistical significance required for a fluctuation 
to be detected at a given confidence level.

\begin{question}
What significance is required to claim the
detection of any one of the 11 possible
absorption lines, i.e., to detect the WHIM towards \pg?
\end{question}

In this case, the analyst would be satisfied with a detection at any redshift.
With this information, the number of resolution elements is needed. As 
discussed in the previous section,
there were a total of twelve possible resolution elements (six for \ovii\
and six for \oviii). Since one
red-shifted line corresponded to another $z=0$ line (which
was not interesting for this search), the total
number of $z > 0$ lines searched is reduced to $N=11$. Notice that the redshift
of the absorption lines was fixed at values set by independent
data (in this case, from \fuv\ observations). Therefore each red-shifted
absorption line only had one opportunity to be detected, at the exact
wavelength specified by the rest energy and the redshift.

Next, the level of significance of the to--be--detected 
feature needs to be determined. This means deciding on a 
cumulative null hypothesis probability $P$ and the associated
individual probability $p$. If one desires an overall 
probability of
detection of 99\% (or, in terms of equivalent Gaussian
standard deviations, $\sim 2.6 \sigma$), then the null hypothesis
probability is $P=0.01$. 
For $N=11$ tries, this means that any individual feature will require
an associated confidence according to Table~\ref{tab:PpN}
(approximating to $N=10$)
of $p=0.001$  (in terms of equivalent Gaussian
standard deviations, $\sim 3.3 \sigma$).

If one chooses a confidence level corresponding to a 
$3 \sigma$ fluctuation ($P=0.003$ or 99.7\% confidence level),
then the individual level of significance would have to rise to $p \simeq 0.0003$
for 10 resolution elements, corresponding to the requirement
of an individual $\sim 3.6 \sigma$ Gaussian fluctuation. The choice of
confidence level is largely a matter of personal preference and best practices
in the field of study. In the following the $3 \sigma$ threshold of detection
is used.

{\blue \begin{answer}
 An  individual $\geq 3.6 \sigma$  feature 
can be claimed as a detection at the 
$3 \sigma$ or 99.7\% level, for the given number of tries (N=11).
\end{answer}}
Therefore any feature detected at an individual significance 
that is lower than the desired $p$ 
(in this example, $3.6\sigma$), should be regarded as
chance fluctuations, at the 99.7\% confidence level.

The next point in the analysis is the determination of the presence 
of significant features.

\begin{question}
Are there any statistically significant detections of 
the 11 possible absorption lines in \cite{bonamente2016}?
\end{question}

For this purpose one needs the statistical significance
of each absorption feature. In \cite{bonamente2016}, these numbers 
were reported in their Table~3 (reproduced in Table~\ref{tab:bonamente2016}), 
along with the 
signal--to--noise ratio of each feature,
calculated using the $\Delta \chi^2_{min}$ method described in Sec.~\ref{sec:probabilityModels}.
In that analysis, two features stood out:
an emission feature at
an individual significance level of $3.5\sigma$ (or $p=0.0005$) and an 
absorption feature
at a significance of $5.2\sigma$, the latter at a redshift of $z=0.0911$.  
The absorption feature is significant at the cumulative $P$ level of $3 \sigma$,
while the emission feature drops just below the threshold (which would require
an individual detection $\geq 3.6 \sigma$).

{\blue
\begin{answer}
The $z=0.0911$ absorption
feature is the only feature detected at a statistical significance $\geq 99.7$\%.
\end{answer}}
Although only indirectly related to the question at hand,
it is useful to also discuss the possible emission feature.
The availability of $N=11$ tries means that there
is a cumulative probability $P \simeq 0.005$ or 0.5\% of having one such fluctuation
by chance. This level of probability corresponds to an equivalent
confidence level of $2.8 \sigma$. In other words, the emission
feature falls below the $3 \sigma$ threshold of probability, once 
the number of possible resolution elements or tries is accounted for.
This is an indication that this feature may in fact be a random
fluctuation, in line with the observation
that there is no immediate plausible explanation
for its presence, as also discussed in \cite{bonamente2016}.

It is useful do discuss whether the approximate $p N$ method
of Section~\ref{sec:approximation} is applicable to the detected absorption line.
The calculation of
the cumulative significance for the absorption feature in the original
publication used the approximate $p N$ method
and resulted in an estimated cumulative significance
level of $\sim 4.6 \sigma$.
Following the discussion of Section~\ref{sec:approximation}, 
the validity of the $p N$ approximation can be easily
verified. For an individual $5.2\sigma$ detection,
$p\simeq 0.0000002$ (Table~\ref{tab:gaussianIntegral}).
Therefore, for $N \simeq 10$, it is clear that $pN \ll 2$, and the $pN$ approximation is
guaranteed to give an accurate estimate of the cumulative probability $P$.
Under these circumstances, there is no need to use the full calculation
according to Equation~\ref{eq:p}, and the $pN$ approximation will suffice.

The arbitrariness in the choice of the confidence level makes it such
many analysts choose a probability level \emph{a posteriori}, i.e.,
after having calculated the individual significances of any fluctuations
and the number of resolution elements.
For example, for these data,
it was a logical choice to set a cumulative detection threshold
of $3 \sigma$ or 99.7\%, such that the `undesirable' emission feature becomes
attributable to chance, while the `desirable' absorption feature remains
significant.  This posterior logic has certainly a place in the overall analysis, given that
an analyst may have additional information of the nature of the fluctuations,
besides its pure statistical significance. 

\subsection{Detection of the specific \oviii\ absorption line at $z=0.0911$}
\label{sec:pg-2}

Another question that can be posed in general terms is: 

\begin{question}
What significance is required to claim
the detection of \oviii\ at $z=0.0911$?
\end{question}

In this case, there is only one resolution element where this absorption 
feature can be, and therefore
the (cumulative) null hypothesis probability associated with this question 
is equal to the significance of the individual detection, since $N=1$.
If a threshold of 3 $\sigma$ or 99.7\% is set, so long as the data have a fluctuation
$\geq 3 \sigma$ at that wavelength, the detection can be claimed as real. 
No allowance for other possibilities of detection has to be made,
since the total number of tries or resolution elements is just $N=1$.
A similar question can be phrased for any of the 11 lines of Table~\ref{tab:bonamente2016}.

In principle, this is a perfectly valid question, with a perfectly valid and simple answer. 
However, the analyst needs to justify \emph{why}
such redshift and absorption line 
was chosen \emph{a priori} to formulate the question, i.e., before
performing the analysis. There must be a compelling
reason why the redshift $z=0.0911$ and the specific line were chosen, i.e.,
a reason that makes that redshift unique, to make this question meaningful. Otherwise, 
the question would appear to be unmotivated, or rather motivated
only by the desire to increase the significance of its detection by ignoring
all possible opportunities of detection of equivalent features at different redshifts,
as in Section~\ref{sec:pg-1}.

{\blue
\begin{answer} A 3~$\sigma$ significance for the line is required for a 3~$\sigma$
detection, assuming just one resolution element in the search. 
\end{answer} }

A follow--up question is applicable to the specific
situation of \cite{bonamente2016}:
\begin{question}
What is the significance of detection of the $z=0.0911$ \oviii\ absorption
feature of \cite{bonamente2016}?
\end{question}
In the case of \cite{bonamente2016}, there was no compelling reason indicating
that the $z=0.0911$ redshift was \emph{special}. 
In other words, the analysis of the source \pg\ and of foreground structures 
did not indicate any reasons why that redshift would be a unique candidate for
an \oviii\ absorption line from the WHIM.
This means that it would be appropriate
to allow for $N=11$ opportunities to detect any absorption lines, as 
discussed in Sec.~\ref{sec:pg-1}.
Instead of reporting the significance of detection of the $z=0.0911$ \oviii\ 
absorption line at its individual level of significance of $5.2\sigma$
(or a null hypothesis probability of $p \simeq 0.0000002$), as claimed
in the abstract of that paper, it would be more accurate to lower
the reported significance of detection according to

\[ P = p N \simeq 0.000002 \]

which is approximately the probability corresponding to a $\sim$ 4.7$\sigma$
deviation, according to Table~\ref{tab:gaussianIntegral}.
This level of significance
remains very high, but there may be other situations where a marginal
detection becomes not significant when the proper number of tries 
is accounted. 

{\blue
\begin{answer} Accounting for $N=11$ resolution elements in the search,
the line is significant at the $\sim 4.7 \sigma$ level ($p\simeq 2\times 10^{-5}$).
\end{answer}
}

\begin{table}
\caption{Values $A$ of two--sided integral of standard Gaussian within $\pm \sigma$ of mean
 and corresponding null hypothesis probability $p=1-A$. Values are 
obtained using \texttt{python} math libraries and are reported with 
eight digits of decimal precision.} 
\label{tab:gaussianIntegral}
\begin{tabular}{llllll}
\hline
\hline
$\sigma$&	$A$		&	$p$		&	$\sigma$&       $A$          &       $p$ 		\\
\hline
0	&	0		&	1		&	3	&	0.9973002	&	0.0026998	\\
0.1	&	0.07965567	&	0.92034433	&	3.1	&	0.99806479	&	0.00193521	\\
0.2	&	0.15851942	&	0.84148058	&	3.2	&	0.99862572	&	0.00137428	\\
0.3	&	0.23582284	&	0.76417716	&	3.3	&	0.99903315	&	0.00096685	\\
0.4	&	0.31084348	&	0.68915652	&	3.4	&	0.99932614	&	0.00067386	\\
0.5	&	0.38292492	&	0.61707508	&	3.5	&	0.99953474	&	0.00046526	\\
0.6	&	0.45149376	&	0.54850624	&	3.6	&	0.99968178	&	0.00031822	\\
0.7	&	0.5160727	&	0.4839273	&	3.7	&	0.9997844	&	0.0002156	\\
0.8	&	0.5762892	&	0.4237108	&	3.8	&	0.9998553	&	0.0001447	\\
0.9	&	0.63187975	&	0.36812025	&	3.9	&	0.99990381	&	0.00009619	\\
1	&	0.68268949	&	0.31731051	&	4	&	0.99993666	&	0.00006334	\\
1.1	&	0.72866788	&	0.27133212	&	4.1	&	0.99995868	&	0.00004132	\\
1.2	&	0.76986066	&	0.23013934	&	4.2	&	0.99997331	&	0.00002669	\\
1.3	&	0.80639903	&	0.19360097	&	4.3	&	0.99998292	&	0.00001708	\\
1.4	&	0.83848668	&	0.16151332	&	4.4	&	0.99998917	&	0.00001083	\\
1.5	&	0.8663856	&	0.1336144	&	4.5	&	0.9999932	&	0.0000068	\\
1.6	&	0.89040142	&	0.10959858	&	4.6	&	0.99999578	&	0.00000422	\\
1.7	&	0.91086907	&	0.08913093	&	4.7	&	0.9999974	&	0.0000026	\\
1.8	&	0.92813936	&	0.07186064	&	4.8	&	0.99999841	&	0.00000159	\\
1.9	&	0.94256688	&	0.05743312	&	4.9	&	0.99999904	&	0.00000096	\\
2	&	0.95449974	&	0.04550026	&	5	&	0.99999943	&	0.00000057	\\
2.1	&	0.96427116	&	0.03572884	&	5.1	&	0.99999966	&	0.00000034	\\
2.2	&	0.9721931	&	0.0278069	&	5.2	&	0.9999998	&	0.0000002	\\
2.3	&	0.97855178	&	0.02144822	&	5.3	&	0.99999988	&	0.00000012	\\
2.4	&	0.98360493	&	0.01639507	&	5.4	&	0.99999993	&	0.00000007	\\
2.5	&	0.98758067	&	0.01241933	&	5.5	&	0.99999996	&	0.00000004	\\
2.6	&	0.99067762	&	0.00932238	&	5.6	&	0.99999998	&	0.00000002	\\
2.7	&	0.99306605	&	0.00693395	&	5.7	&	0.99999999	&	0.00000001	\\
2.8	&	0.99488974	&	0.00511026	&	5.8	&	0.99999999	&	0.00000001	\\
2.9	&	0.99626837	&	0.00373163	&	5.9	&	1		&	0		\\
\hline
\hline
\end{tabular}
\end{table}

\subsection{Re--analysis of probabilities allowing an error in the redshift}
\label{sec:z}

This case study also lends itself to other considerations
that are worth discussing in more detail. 
The search for absorption lines was initially performed at
the exact redshifts indicated by the FUV data of \cite{tilton2012}. 
At these fixed redshifts, the significance of the only
absorption line detected at an individual significance $\geq 3 \sigma$  
was \oviii\ at the FUV redshift of $z=0.0928$ 
(with negligible error), detected with a significance of $3.7 \sigma$.
For this line,
the analysis was refined by allowing
the redshift to vary around its nominal FUV value.
Reason for this refinement in the analysis is that several sources of error
may give rise to a slight change in wavelength between the FUV and X--ray observations.
The significance rose to $5.2 \sigma$ when a free redshift was allowed,
and measured in an amount of $z=0.0911\pm0.0004\pm0.0005$ \citep[as reported in][]{bonamente2016}. 
This is the redshift used earlier in this section.

The double error--bar notation is used to identify two sources of error,
the first due to so--called statistical sources (primarily counting statistics)
and the second due to systematic sources (primarily the accuracy 
of the wavelength scale in the detector, as also discussed in \cite{bonamente2016}).
First of all, it is not clear how the two sources of errors should be added,
namely whether in quadrature or linearly, or in a different manner.
A quadrature addition would be appropriate if the variable of
interest (the fluctuation $\Delta K$) can be thought of as a sum
$\Delta K = K_1 + K_2$,
where $K_1$ is the measurement of the fluctuation and $K_2$ is
a random variable with zero mean, representing the systematic error.
Further, the systematic error must be uncorrelated to the measurement
for the quadrature addition to apply.
This is likely to be the case, since it is assumed that 
the same systematic error applies regardless of the
value of the measurement $K_1$. 
A quadrature addition of the errors would then be summarized as
$z=0.0911\pm0.0007$, showing that this X--ray redshift is
within 3 standard deviations of the FUV redshift $z=0.0928$,
as the z--score of the difference is -2.4. 
This criterion can be used to say that the two redshifts are
statistically consistent, at the $3 \sigma$ level.
This statistical consistency can be used to say that 
the proper value for the measurement of the fluctuation
has a significance of $5.2\sigma$, corresponding to the
adjusted redshift. This was the argument used in \cite{bonamente2016},
based on the fact that (a) the initial search at the fixed FUV redshift 
yielded a $\geq 3 \sigma$ fluctuation and that (b) knowledge of
the detectors allows a slight adjustment.

It is necessary to investigate the effect of a redshift error in more detail.
To begin with:
\begin{question}
How is the total number of tries $N$ affected by an uncertainty  in redshift
in the search for the WHIM absorption lines?
\end{question}

This allowance for an `adjustment' of the redshift 
requires a correction to the total number of
tries or resolution elements $N$ available in the search.
If one allows a range of $\pm 3 \sigma$ around the nominal
FUV redshift (where the standard deviation $\sigma$ is from
the X--ray measurements themselves), then
each line may have more than just one resolution
element where it can be detected.
It was assumed for simplicity that there is a redshift range of $\sigma_z = \pm (3 \times 0.0007)$
where each line can be detected, around the nominal redshift 
and therefore the nominal wavelength.
This redshift range corresponds to a wavelength range of

\[ \Delta \lambda = 2 \lambda_0 \sigma_z = 91~\text{m\AA} \]

for the $\lambda_0=21.6$~\AA\ \ovii\ line (a similar
value apply for the \oviii\ lines), where $\sigma_z=0.0021$, and the factor of 2
is used to account for positive and negative adjustments. This redshift
range (91 \text{m\AA}) is approximately twice the size of the resolution 
element ($\sigma_{\lambda} = 50$~m\AA), and therefore 
it is reasonable to assume that there are \emph{two} resolution
elements where each line has an opportunity to be detected, not just one.
Following this argument, one can now say that there was a total of $N=22$
(and not $N=11$ as assumed so far)
resolution elements in the search -- surely, this same procedure
would have been applied to any of the 11 possible lines, had they been detected.

{\blue 
\begin{answer}
Using a redshift range of $\sim 90~m\AA$ around each line
with a resolution of 50$~m\AA$, the number of resolution
elements doubles to $N=22$.
\end{answer}
}

Finally we can restate an earlier question, accounting for errors in the
redshift:
\begin{question}
What is the significance of detection of the $z=0.0911$ \oviii\ absorption feature
of \cite{bonamente2016}, accounting for errors in the redshift?
\end{question}

For $N=22$, an individual detection at the 5.2$\sigma$ level 
($p \simeq 0.0000002$) 
corresponds to a cumulative null hypothesis probability of approximately 
$P= pN \simeq 0.000004$, where one can
use the approximate $pN$ method since $pN$ is very small. 
 Such value of $pN$ corresponds
to a $\simeq 4.6 \sigma$ fluctuation, i.e., 
an individual 5.2$\sigma$ detection remains very significant
also for the increased value of $N=22$. 
Although in this  case study
this correction did not affect the result, in general
it is important to allow errors in the redshifts
to calculate the correct $N$ and the corresponding cumulative probability $P$.
Marginal detections may become not significant when the 
proper number of tries is accounted.

{\blue
\begin{answer}
Allowing for the redshift uncertainty, the significance is reduced
from 4.7 $\sigma$ to 4.6 $\sigma$.
\end{answer}
}

\subsection{Probability of detection of both \ovii\ K$\alpha$ and \ovii\ K$\beta$
or other pairs of lines}

Having established that the \oviii\ K$\alpha$ line is statistically significant,
\cite{bonamente2016} proceeded to seek the presence of the associated K$\beta$ line,
caused by a related atomic transition in the same ion.
If \oviii\ K$\alpha$ is detected,  the weaker K$\beta$ line must also present.
The K$\beta$ line is located at 16.00~\AA\ in the rest frame, therefore at a fixed distance
from the  K$\alpha$ line, and with a prescribed ratio of intensities.

\begin{question}
Is the \oviii\ K$\beta$ absorption line at $z=0.0911$ in \cite{bonamente2016}
statistically significant?
\end{question}

The method to detect the \oviii\ K$\beta$ was to add another one--parameter model
for this line at the prescribed wavelength, corresponding to the \emph{same} redshift as
for the K$\alpha$ line. The parameter is proportional to the intensity of the line, or number of
counts in the line. At the wavelength of the \oviii\ K$\beta$ line, the spectrum had
a deficit of counts corresponding to a negative fluctuation 
with a flux of $-13.6 \pm 5.6$ (in same units as Table~\ref{tab:bonamente2016}), 
therefore with 2.4$\sigma$ significance.

The analysis of \cite{bonamente2016} can be used to address the question at hand.
Given that the search for \oviii\ K$\beta$ commenced \emph{after} the detection of K$\alpha$
was established, this new search surely has only one possibility of detection ($N=1$). Therefore
the 2.4$\sigma$ significance can be taken at face value, and say that the K$\beta$
line is \emph{not} detected at the 3$\sigma$ (or 99.7\%) level, but just at the
2.4$\sigma$ or 98.4\% level. There is a 1.6\% chance that this detection is spurious.

{\blue
\begin{answer}
The \oviii\ K$\beta$ absorption line is not statistically significant, at the $3 \sigma$
level.
\end{answer}
}

This independent modeling of the K$\alpha$ and K$\beta$ lines, however,
 misses the fact that the relative intensity of the two lines is prescribed.

\begin{question}
Is the combined detection of \oviii\ K$\alpha$ and K$\beta$ at $z=0.0911$ in \cite{bonamente2016}
statistically significant?
\end{question}
In \cite{bonamente2016} no effort was made to describe the joint probability of 
detecting both \oviii\ K$\alpha$ and K$\beta$. A simple way to find this probability would 
have been to link the intensities of the two lines and use a one--parameter model for the
joint $\Delta K$, as explained in case (a) of Section~\ref{sec:method2-app}. 
This is certainly the most direct way to assess the significance
of two lines that are at a fixed distance and with a prescribed
ratio of intensities.

Based on the data of \cite{bonamente2016}, the closest one can come to provide a quantitative
answer to this question is to combine the individual K$\alpha$ and K$\beta$ detections,
for a total flux deficit of $-45.0 \pm 8.2$ or a 5.5$\sigma$ detection.
The usual error propagation formula
$\sigma^2_{1,2} = \sigma^2_1 + \sigma^2_2$
was used to add the errors \citep[e.g.][]{bevington2003, bonamente2017book}.
Given that \oviii\ K$\alpha$ was already detected at any reasonable
level of significance, clearly the combination of the two is also significant.

{\blue
\begin{answer}
Yes, because the significance is driven by the \oviii\ K$\alpha$ line.
\end{answer}
}
An independent model for two lines of interest is justified
when the intensities of two lines are not linked, as explained in
case (b) of Section~\ref{sec:method2-app}. For example:

\begin{question}
Is oxygen (either \ovii\ or \oviii) significantly 
detected at $z=0.041$ in the spectrum of \pg? 
\end{question}

The relative intensity of \ovii\ and \oviii\ 
can have any value, 
since it is a function of the unknown temperature. Therefore independent
models for the two lines are necessary.
According
to Table~\ref{tab:bonamente2016}, there is a 1.1$\sigma$ detection of \ovii\
and a 1.0 $\sigma$ detection of \oviii, i.e., neither line is significantly detected
individually.
It is assumed that there is an a priori reason to seek oxygen at that 
specific redshift, so that the redshift for the search is fixed. 
Therefore one seeks fluctuations in two possible resolution elements ($N=2$),
at the two wavelengths of redshifted \ovii\ and \oviii.
Following this assumption,
the question can be re--stated as: what is the chance probability
of two $1 \sigma$ fluctuations (or $p=0.32$) out of two possible tries?
According to Table~\ref{tab:PpN-2}, there is a 10\% chance probability
of two $1 \sigma$ fluctuations out of $N=2$ tries. Therefore one could conclude
that oxygen is detected at the 90\% confidence level. Usually this confidence level is 
considered too low to claim a conclusive detection. The main point of this 
example is to illustrate how to use the model of Section~\ref{sec:model2} 
(and Tables~\ref{tab:PpN-2}-\ref{tab:PpN-3}) for multiple fluctuations.

{\blue
\begin{answer}
No, the individual 1.1 and 1.0 $\sigma$ detections do not combine to
an overall significant detection of oxygen at that redshift.
\end{answer}
}

In the absence of this method for multiple fluctuations, 
one could have answered the question by combining the signals
at the two wavelengths.
The combined \ovii\ and \oviii\ signal at $z=0.041$ is calculated 
as $\Delta K_{1,2} = -14.5 \pm 9.9$ using the numbers 
from Table~\ref{tab:bonamente2016}.
This corresponds to a $\sim 1.5 \sigma$ detection with a
null hypothesis probability of $\sim 13$\%. 

Notice how the answer based on the method described in Section~\ref{sec:model2}
differs from that based on the approximate method of combining the two
independent fluctuations. 
The differences between the two methods are even more significant 
for a larger number of fluctuations. Consider as an example three identical
$1 \sigma$ fluctuations that can occur in $N=3$ resolution elements. 
According to a simple combination of the signals, one would obtain 
a $\sqrt{3}=1.7 \sigma$ fluctuation, with a null hypothesis of 9\%.
However, the calculation based on the method of Section~\ref{sec:model2}
results in a null hypothesis probability of just 3\%, according to Table~\ref{tab:PpN-3}. The method
for multiple fluctuations described in this paper can therefore be 
quite useful for assessing the combined  statistical significance of the
combination of low--significance features.

\section{Conclusions}
This paper addressed  the probability of 
occurrence of unresolved fluctuations, such 
as absorption and emission lines 
in astronomical spectra, with the
goal of determining whether observed features are
statistically significant or simply due to chance fluctuations based
on the noise in the data. 
The proposed method generalizes and
significantly expands the current best--practices methods of analysis used 
in the field.

The statistical modeling
is based on a binomial probability distribution for the fluctuations.
The calculation of chance probability of fluctuations depend
on the number of independent resolution elements and
the probability level for the fluctuations. This note provides tables
and formulas
that will aid the analyst in deciding what is a reasonable
confidence level to accept detected fluctuations.

The paper is  complemented
by a case study based on the X--ray data of \cite{bonamente2016}.
The case study highlights the application of the model 
and other practical aspects in the search for unresolved fluctuations
in astronomical spectra, such as the use of \emph{a priori}
assumptions on the location of fluctuations and their impact on the
probability of detection.

\bibliographystyle{apj}

\end{document}